\documentclass[12pt]{article}
\usepackage{latexsym}
\def\hybrid{\topmargin -20pt    \oddsidemargin 0pt
        \headheight 0pt \headsep 0pt
        \textwidth 6.5in        
        \textheight 9in         
        \marginparwidth .875in
        \parskip 5pt plus 1pt   \jot = 1.5ex}

\hybrid

\renewcommand{\arraystretch}{1.5}

\makeatletter
\@addtoreset{equation}{section}
\makeatother

\newcommand{\color}[6]{}

\newcommand{\cA}{{\cal A}}

\newcommand{\cD}{{\cal D}}

\newcommand{\cF}{{\cal F}}
\newcommand{\cL}{{\cal L}}

\newcommand{\cN}{{\cal N}}
\newcommand{\cO}{{\cal O}}
\newcommand{\cP}{{\cal P}}

\newcommand{\cV}{{\cal V}}
\newcommand{\cW}{{\cal W}}

\newcommand{\hf}{\frac12}
\newcommand{\qt}{\frac14}
\newcommand{\bea}{\begin{eqnarray}}
\newcommand{\eea}{\end{eqnarray}}
\newcommand{\be}{\begin{equation}}
\newcommand{\ee}{\end{equation}}
\newcommand{\bt}{\begin{tabular}}
\newcommand{\et}{\end{tabular}}
\newcommand{\ba}{\begin{array}}
\newcommand{\ea}{\end{array}}

\newcommand{\diag}{{\rm diag}}

\newcommand{\Mpl}{M_{\rm Pl}}
\def\IB{\relax{\rm I\kern-.18em B}}
\def\ID{\relax{\rm I\kern-.18em D}}
\def\IE{\relax{\rm I\kern-.18em E}}
\def\IF{\relax{\rm I\kern-.18em F}}
\def\IH{\relax{\rm I\kern-.18em H}}
\def\II{\relax{\rm I\kern-.18em I}}
\def\IK{\relax{\rm I\kern-.18em K}}
\def\IL{\relax{\rm I\kern-.18em L}}
\def\IM{\relax{\rm I\kern-.18em M}}
\def\IN{\relax{\rm I\kern-.18em N}}
\def\IP{\relax{\rm I\kern-.18em P}}
\def\IR{\relax{\rm I\kern-.18em R}}
\def\IT{\relax{\rm I\kern-.42em T}}
\def\IZ{\relax{\hbox{\raisebox{.38ex}
    {\scriptsize\bfseries\slshape /}\kern-.40em\_\kern-.28em\rm Z}}}
\def\Iz{\relax{\hbox{\raisebox{.38ex}
    {\tiny\bfseries\slshape /}\kern-.25em\raisebox{.65ex}
    {\tiny\bfseries\slshape /}\kern-.43em\_\kern-.26em\rm Z}}}
\def\inbar{\vrule height1.5ex width.8pt depth-0.2pt}
\def\inbarhi{\vrule height1.55ex width.5pt depth-.85ex}
\def\inbarlo{\vrule height.8ex width.5pt depth0ex}
\def\IC{\relax{\rm C\kern-.48em \inbar\kern.48em}}
\def\IO{\relax{\rm O\kern-.56em \inbar\kern.56em}}
\def\IQ{\relax{\rm Q\kern-.56em \inbar\kern.56em}}
\def\IS{\relax{\rm S\kern-.37em \inbarhi\kern.08em\inbarlo\kern.29em}}

\def \one{\relax{\rm 1\kern-.26em I}}

\def \soll={\stackrel{!}{=}}

\def \bfm#1{\mbox{\boldmath$#1$}}

\def \Mpl{M_{\rm Pl}}

\def \eps{\epsilon}

\def \barfill{\leaders\hrule height 0.1 true pt\hfill}
\def \overbar#1{\vbox{\ialign{##\crcr\barfill\crcr\noalign{\kern 1pt
                                      \nointerlineskip}$\hfil{#1}\hfil$\crcr}}}
\def \scriptbar#1{{\vbox{\ialign{##\crcr\thinspace\barfill\thinspace\crcr
    \noalign{\kern 0.8pt\nointerlineskip}$\hfil{\scriptstyle #1}\hfil$\crcr}}}}

\newlength{\oldindent}
\newlength{\quadlength} 
\newlength{\abstand} \newlength{\breite}

\def \Nucl#1{{\em Nucl.~Phys.}\ {\bf B#1}}

\def \PhysR#1{{\em Phys.~Rev.}\ {\bf D#1}}
\def \PhysRL#1{{\em Phys.~Rev.~Lett.}\ {\bf #1}}

\def \PhysL#1{{\em Phys.~Lett.}\ {\bf #1B}}

\def \vm{v^m_\xi}
\def \vn{v^n_\xi}
\def \vk{v^k_\xi}
\def \Vm{V^m_\eta}

\def \pseudo{pseu\-do\discretionary{-}{}{-}su\-per\-sym\-me\-try}

\renewcommand{\thefootnote}{\fnsymbol{footnote}}

\begin{document}

\begin{titlepage}
\begin{center}

\rightline{}
\rightline{SLAC-PUB-9234}
\rightline{hep-th/0205300}

\vskip .6in
{\LARGE \bf Couplings in Pseudo-Supersymmetry}
\vskip .8in

{\bf Matthias Klein\footnote{E-mail: mklein@slac.stanford.edu}}
\vskip 0.8cm
{\em SLAC, Stanford University, Stanford, CA 94309, USA.}

\end{center}

\vskip 1.5cm

\begin{center} {\bf ABSTRACT } \end{center}
We analyze theories in which a supersymmetric sector is coupled to
a supersymmetry-breaking sector described by a non-linear realization.
We show how to consistently couple $\cN=1$ supersymmetric matter
to non-supersymmetric matter in such a way that all interactions are
invariant under non-linear supersymmetry transformations. We extend
this formalism to couple $\cN=2$ supersymmetric matter to $\cN=1$ superfields
that lack $\cN=2$ partners but transform in a non-linear representation of
the $\cN=2$ algebra. In particular, we show how to couple an $\cN=2$ vector 
to $\cN=1$ chiral fields in a consistent way.
This has important applications to effective field theories describing 
the interactions of D-brane world-volume fields with bulk fields.
We apply our method to study systems where different sectors break
different halves of supersymmetry, which appear naturally in models
of intersecting branes.

\vfill

May 2002
\end{titlepage}

\newpage

\setcounter{page}{1} \pagestyle{plain}
\renewcommand{\thefootnote}{\arabic{footnote}}
\setcounter{footnote}{0}


\section{Introduction}
With the advent of D-branes \cite{Polch}, string theory has provided new 
and exciting mechanisms of supersymmetry breaking. Traditionally, most
semi-realistic string models were supersymmetric at the string scale and
supersymmetry was broken by a field theory effect. The auxiliary field
of some supermultiplet would get a vacuum expectation value in the effective
supergravity theory. This gave rise to mass splittings for the supermultiplets
through the super-Higgs effect. Since the discovery of D-branes, it became
clear that there are phenomenologically interesting models where 
supersymmetry is broken by stringy effects (e.g., \cite{ADS,AU,AIQ,AADDS}).
Parallel D-branes break half of the supersymmetry that is present in the
bulk. The fields arising from the open string excitations, which are confined
to the world-volume of the D-branes, only fill multiplets of the smaller
supersymmetry algebra. Part of the supersymmetry is explicitly broken on 
the D-branes, since the world-volume fields lack the corresponding
superpartners. Supersymmetry can be completely broken by adding anti-D-branes,
which preserve the other half of the bulk supersymmetry.
Thus supersymmetry is broken in a non-local way in such models. Each sector 
taken separately preserves part of the supersymmetry.
A very similar situation arises in models containing stacks of D-branes
at angles that intersect each other (e.g., \cite{BGKL,quasi}). There is
an extended supersymmetry on each stack of D-branes, but only a fraction
of this supersymmetry is preserved at each intersection. Supersymmetry is
completely broken in models where different intersections break different
fractions of supersymmetry. 

The field theory of such models is interesting in its own right.
Supersymmetry is broken explicitly but non-locally. As a consequence,
there are no mass splittings and no quadratic divergences at one-loop
\cite{quasi}. In the present article, we would like to study an effective
field theory description of this supersymmetry breaking mechanism.

To determine the couplings of the bulk fields to the boundary fields, it
is important to note that the part of supersymmetry that is broken on the
D-branes is still non-linearly realized. This statement has not been
rigorously proven but there is much evidence in favor of it. For concreteness,
consider a single D3-brane in flat ten-dimensional space. From the
space-time point of view the translational invariance transverse to the
brane is spontaneously broken in the sense that the brane is located at
some definite position but all possible positions have degenerate potentials
\cite{string}. Similarly, half of the supersymmetries are spontaneously
broken. From the world-volume theory point of view the breaking seems to
be explicit since the world-volume fields lack the superpartners that
would correspond to the broken supersymmetries. However, the broken 
symmetries are non-linearly realized and it is easy to identify the
associated Goldstone fields. In the world-volume theory, position in
transverse space is parameterized by the expectation value of the six
scalar fields inside the $\cN=4$ vector multiplet. These scalars having
no potential are the Goldstone bosons of broken translational invariance.
Interestingly, the $\cN=4$ vector multiplet contains just the right number
of fermions to provide four goldstinos corresponding to the breaking
of the original $\cN=8$ supersymmetry down to $\cN=4$.
Let us now compactify four of the transverse dimensions on an orbifold that
preserves half of the supersymmetry and place the D3-brane at an
orbifold singularity. This projects out half of the fields on the D3-brane
world-volume and leads to an $\cN=2$ $U(1)$ gauge theory. The two scalars
inside the $\cN=2$ vector are the Goldstone bosons corresponding to broken
translational invariance in the two directions where the brane can still
be moved. Again the goldstinos corresponding to the broken supersymmetries
are the superpartners of the Goldstone bosons of translation symmetry.
If we compactify all six transverse dimensions on an orbifold that preserves 
only a quarter of bulk supersymmetry and place the D3-brane at an orbifold
singularity, then there is no modulus left that would correspond to the
motion of the brane in transverse space. Indeed, there is no scalar in 
the $\cN=1$ vector that survives the orbifold projection on the brane.

In a beautiful paper \cite{BG}, Bagger and Galperin have analyzed the
goldstino couplings that are necessary to have non-linearly realized
$\cN=2$ supersymmetry in a manifestly $\cN=1$ invariant $U(1)$ gauge
theory. Amazingly, the assumption that the gaugino is the goldstino
of partially broken $\cN=2$ supersymmetry implies a supersymmetric
generalization of the Born-Infeld action for the $U(1)$ vector multiplet.
This is further evidence for the conjecture that the supersymmetries that 
are broken on the D-branes are really non-linearly realized.
In this article, we use the formalism of partially broken supersymmetry
developed in \cite{BG} to determine the bulk-to-boundary couplings, that
is the couplings of fields on which $\cN=2$ supersymmetry is linearly
realized to fields that form non-linear realizations of $\cN=2$ supersymmetry.%
\footnote{This work is related to a problem discussed in \cite{BFKQ}. 
It was motivated by \cite{MP} where the couplings of an
$\cN=2$ bulk vector to $\cN=1$ boundary chiral fields were determined. A
crucial difference is that in \cite{MP}, half of the components of the bulk
vector were projected out on the boundary, whereas in the present work, all
components of bulk fields survive on the boundary.}
We apply our results to situations where different sectors break different
halves of supersymmetry, but all sectors realize supersymmetry (at least)
non-linearly. We call this scenario {\em \pseudo}.

Additional evidence for non-linearly realized supersymmetry on D-branes
was pointed out by the authors of \cite{DM} (see also \cite{PR}). They 
argued that consistent gravitino couplings are very constrained and that 
it is hard to imagine how these constraints can be satisfied without 
supersymmetry. For a non-supersymmetric type I string model they showed 
that supersymmetry broken on the D-branes is indeed non-linearly realized. 
One of the gauginos of the world-volume theory is a gauge singlet and has 
just the right couplings to be the Goldstone fermion of broken supersymmetry.
In contrast to models of spontaneously broken supersymmetry, however, no
description of these string models in terms of linear supersymmetry broken
by some super-Higgs effect is known. The scale of supersymmetry breaking
is not a tunable vacuum expectation value of some field but it is tied to 
the string scale. The non-linear goldstino couplings have been determined 
by performing a string computation in \cite{ABL}.

The paper is organized as follows. In the next section we review
non-linear realizations of supersymmetry. In section 3, it is shown 
how to couple $\cN=1$ sectors to $\cN=0$ sectors in such a way that all 
interactions are invariant under either linear or non-linear supersymmetry.
We then review and generalize the formalism of non-linear realizations
applied to partially broken $\cN=2$ supersymmetry. This formalism together
with the method to couple sectors of non-linear supersymmetry to sectors
of linear supersymmetry are used in section 5 to determine the couplings 
of $\cN=2$ multiplets to $\cN=1$ matter. In section 6, we apply these
results to a toy model of \pseudo. Some open questions
are outlined in the outlook. Finally, our notation is explained in the
appendix.

\section{Review of non-linearly realized supersymmetry}
Supersymmetry is certainly broken at low energies. It is therefore not
very surprising that the non-linear realization of supersymmetry was
analyzed \cite{AV,AV_tmp,Pash,VS} even before the importance of linear 
supersymmetry \cite{WZ} was realized. Let us briefly review the 
formalism developed in \cite{AV,AV_tmp,Pash} in our notation.

The simplest model of non-linearly realized supersymmetry contains just 
one fermion, the goldstino $\lambda_g$. The supersymmetry variation
of the goldstino is \cite{AV}
\be  \label{gstino_transf}
\delta_\xi\lambda_g={1\over\kappa}\xi-\kappa\,\vm\partial_m\lambda_g,
\qquad{\rm where}\quad 
\vm=i\lambda_g\sigma^m\bar\xi-i\xi\sigma^m\bar\lambda_g.
\ee
The constant $\kappa$ has mass dimension $-2$. It can be interpreted as 
the scale of supersymmetry breaking. We use the spinor conventions of
Wess and Bagger \cite{WB}.

This transformation realizes the supersymmetry algebra \cite{AV_tmp},
\be  \label{gstino_algebra}
[\delta_\eta,\delta_\xi]\,\lambda_g=
       -2i\,(\eta\sigma^m\bar\xi-\xi\sigma^m\bar\eta)\,\partial_m\lambda_g.
\ee

To construct an invariant action, we define \cite{AV}
\be  \label{omega_def}
\omega_m^{\ n}=\delta_m^{\ n}
                 -i\kappa^2\,\partial_m\lambda_g\sigma^n\bar\lambda_g
                 +i\kappa^2\,\lambda_g\sigma^n\partial_m\bar\lambda_g
\ee
and note that 
\be  \label{detom_transf}
\delta_\xi\det(\omega)=-\kappa\,\partial_m\left(\vm\det(\omega)\right).
\ee
This shows that \cite{AV}
\bea  \label{gstino_action}
S_{\rm goldstino} &= &-{1\over2\kappa^2}\int d^4x\,\det(\omega) \\
     &= &\int d^4x\left( -{1\over2\kappa^2}
            -{i\over2}(\lambda_g\sigma^m\partial_m\bar\lambda_g
                       -\partial_m\lambda_g\sigma^m\bar\lambda_g)
            \ + \ \cO(\kappa^2)\right)  \nonumber
\eea
is invariant under the non-linear supersymmetry transformation 
(\ref{gstino_transf}).

The supersymmetry algebra can be realized on any field $f$ --- which may
be a Lorentz scalar, spinor, vector or tensor, but we suppress all
Lorentz indices for simplicity --- by assigning to it the transformation
law \cite{AV_tmp,Pash}
\be  \label{matter_transf}
\delta_\xi f=-\kappa\,\vm\partial_m f.
\ee
It is convenient to introduce covariant derivatives \cite{IK},
\be   \label{cov_deriv}
D_m f = \left(\omega^{-1}\right)_m^{\ n}\partial_n f,
\ee
such that $\delta_\xi(D_m f)=-\kappa\,\vn\partial_n(D_m f)$.
To be able to generalize this to gauge covariant derivatives, we 
assign a transformation law to gauge fields that differs from the
standard non-linear realization (\ref{matter_transf}) but nevertheless
realizes the supersymmetry algebra \cite{CL,CLLW},
\be  \label{gauge_transf}
\delta_\xi A_m=-\kappa\,\vn\partial_nA_m-\kappa\,(\partial_m\vn) A_n.
\ee
This implies that the gauge covariant derivative \cite{CLLW}
\be  \label{gauge_cov_deriv}
\cD_m f\ =\ \left(\omega^{-1}\right)_m^{\ n}
            \left(\partial_n f + iA_n f\right)
\ee
transforms as $\delta_\xi(\cD_m f)=-\kappa\,\vn\partial_n(\cD_m f)$
under the non-linear supersymmetry. Finally, we define \cite{CLLW}
\be  \label{cov_field_strength}
\cF_{mn}=\left(\omega^{-1}\right)_m^{\ k}\left(\omega^{-1}\right)_n^{\ l}
          \left(\partial_kA_l-\partial_lA_k-i[A_k,A_l]\right),
\ee
which transforms as $\delta_\xi\cF_{mn}=-\kappa\,\vk\partial_k\cF_{mn}$.

It is now clear that 
\be  \label{matter_action}
S_{\rm matter}\ =\ \int d^4x\,\det(\omega)\left(
                       -\cD_m f \left(\cD^m f\right)^\dagger
                       -V(f)-\frac14\cF^{mn}\cF_{mn}\right)
\ee
is invariant under (\ref{gstino_transf}), (\ref{matter_transf}),
(\ref{gauge_transf}) for any function $V$.
More generally, any Lorentz invariant Lagrangian $\cL_0$
can be made invariant under the non-linear supersymmetry by coupling
it to the goldstino via
\be   \label{gstino_coupl}
\cL_{\rm susy}\ =\ \det(\omega)\cL'
              \ =\ \cL'
                   +i\kappa^2
                       (\lambda_g\sigma^m\partial_m\bar\lambda_g
                       -\partial_m\lambda_g\sigma^m\bar\lambda_g)
                    \cL'
                   +\cO(\kappa^4),
\ee
where $\cL'$ is obtained from $\cL_0$ by replacing all partial or
gauge covariant derivatives by the supercovariant derivatives defined
above and $F_{mn}$ by $\cF_{mn}$.
It is easy to see that $\cL_{\rm susy}$ transforms into a total derivative,
$\delta_\xi\cL_{\rm susy}=-\kappa\,\partial_m(\vm\cL_{\rm susy})$.

Sometimes it is useful to consider a different non-linear realization
whose action on the goldstino $\tilde\lambda_g$ is defined by \cite{Z,SW}
\be  \label{gstino_chiral}
\delta_\xi\tilde\lambda_g={1\over\kappa}\xi-2i\kappa\,
              \tilde\lambda_g\sigma^m\bar\xi\,\partial_m\tilde\lambda_g.
\ee
This transformation is related to the standard transformation through \
a field redefinition. One can show \cite{IK} that 
$\lambda_g(x^m)\equiv\tilde\lambda_g(x^m+i\kappa^2
                      \lambda_g\sigma^m\bar\lambda_g)$
transforms as in (\ref{gstino_transf}).
We call (\ref{gstino_chiral}) the chiral version of the non-linear
realization (\ref{gstino_transf}) because it involves only left-handed
spinors. 

The action of the chiral non-linear supersymmetry on matter fields is
given by
\be  \label{matter_chiral}
\delta_\xi f= -2i\kappa\,\tilde\lambda_g\sigma^m\bar\xi\,\partial_m f.
\ee

\section{Non-linear realizations from linear realizations}
Consider theories of the form $\cL=\cL_1+\cL_2+\cL_{12}$, where the 
fields in $\cL_1$ have no superpartners and realize supersymmetry
non-linearly, and the fields in $\cL_2$ are in $\cN=1$ 
supermultiplets with a linear realization of supersymmetry. The term
$\cL_{12}$ contains the couplings between the two sectors. We want to
find the restrictions on these couplings arising from the requirement
of supersymmetry. In other words, we would like
to determine how to consistently couple the sector of linearly realized 
supersymmetry to the sector of non-linearly realized supersymmetry.
Our strategy will be to find for each superfield 
$F=(f,\psi_1,\bar\psi_2,\ldots)$ 
a composite field $\hat f$ consisting of the goldstino and the components 
of $F$ which, in the limit $\kappa\to0$, reduces to $f$, the lowest 
component of $F$. This composite field has to be chosen such that the 
non-linear transformation of the goldstino and the linear transformations 
of the components of $F$ induce the standard non-linear realization on 
$\hat f$. The composite field $\hat f$ can be consistently coupled to the 
sector of broken supersymmetry. Thus, if we know how $f$ couples to the
non-supersymmetric matter in the limit $\kappa\to0$, then the
supersymmetric completion of these couplings is obtained by replacing
$f$ by $\hat f$.

\subsection{Coupling a chiral superfield to non-supersymmetric matter}

To clarify what we mean by this, let us discuss a simple example.
Consider a chiral superfield,
\be  \label{Phi_comp}
\Phi(x_L)=\phi(x_L)+\sqrt2\,\theta\psi(x_L)+\theta\theta F(x_L),\qquad
{\rm where}\ \ x_L^m=x^m+i\theta\sigma^m\bar\theta.
\ee
The supersymmetry transformations are
\bea  \label{chiral_transf}
\delta_\xi\phi&=&\sqrt2\,\xi\psi, \nonumber\\
\delta_\xi\psi&=&i\sqrt2\,\sigma^m\bar\xi\partial_m\phi+\sqrt2\,\xi F,\\
\delta_\xi F  &=&i\sqrt2\,\bar\xi\bar\sigma^m\partial_m\psi. \nonumber
\eea
It is straightforward to check that the composite field
\be  \label{phi_hat_L}
\hat\phi_L=\phi-\kappa\sqrt2\,\tilde\lambda_g\psi
             +\kappa^2\,\tilde\lambda_g\tilde\lambda_g F
\ee
transforms precisely according to the chiral non-linear realization
(\ref{matter_chiral}) if the goldstino $\tilde\lambda_g$ transforms
according to (\ref{gstino_chiral}).
Moreover, one can verify that 
\bea  \label{phi_hat}
\hat\phi(x^m) &\equiv &\hat\phi_L(x^m
                  +i\kappa^2\lambda_g\sigma^m\bar\lambda_g)
                  \nonumber\\
              &= &\phi-\kappa\sqrt2\,\lambda_g\psi
                  +\kappa^2\,\lambda_g\lambda_g F
                  +i\kappa^2\lambda_g\sigma^m\bar\lambda_g\partial_m\phi
                  \nonumber\\
              &&  +{i\over\sqrt2}\kappa^3\lambda_g\lambda_g\partial_m\psi
                    \sigma^m\bar\lambda_g
                  +{1\over4}\kappa^4 \lambda_g\lambda_g
                    \bar\lambda_g\bar\lambda_g\Box\phi,
\eea
where all fields in the last two lines are taken at the argument $x^m$,
transforms precisely according to the standard non-linear realization
(\ref{matter_transf}). 
This shows that the components of the superfield $\Phi$ can be 
consistently coupled to sectors of non-linearly realized supersymmetry 
through the combination $\hat\phi$. 

Let us work out the couplings of a bulk dilaton to brane gauge fields.
We consider a model where supersymmetry is completely broken on the 
brane and the only degrees of freedom in the effective theory are the 
gauge fields. The bulk is $\cN=1$ supersymmetric and the dilaton is in
a chiral supermultiplet $\Phi=(\phi,\psi,F)$. Using the above result
and the formalism outlined in the previous section, it is easy to
find the supersymmetric completion of the dilaton coupling term
$(\phi+\phi^\dagger)\,F^{mn}F_{mn}$. By construction,
\bea  \label{nonlin_dil}
S &= &\int d^4x\,\det(\omega)\,(\hat\phi+\hat\phi^\dagger)\,\cF^{mn}\cF_{mn}\\
  &= &\int d^4x\,\left((\phi+\phi^\dagger
                        -\kappa\sqrt2\,(\lambda_g\psi+\bar\lambda_g\bar\psi))
                        \,F^{mn}F_{mn}+\cO(\kappa^2)\right)\nonumber
\eea
is invariant under the non-linear supersymmetry transformations.

\subsection{General formalism}
It is straightforward to generalize the above result for chiral superfields
to arbitrary superfields. We note that (\ref{phi_hat}) is just the usual
superspace expansion of a chiral superfield with the Grassmann variable
$\theta$ replaced by $-\kappa\,\lambda_g$. We will now show that this
prescription is valid for any superfield. That is, from the components
of an arbitrary superfield $F$, one can build a composite field that
transforms as in (\ref{matter_transf}) by replacing 
$\theta\to-\kappa\lambda_g$ in the superspace expansion of $F$.
This result was already discovered long ago \cite{IK}. We will rederive
it here taking a slightly different approach.

Let $F$ be an arbitrary superfield with component expansion
\bea  \label{F_expans}
F &= &e^{\theta Q+\bar\theta\bar Q}f \nonumber\\ 
  &= &f + \theta\psi_1 + \bar\theta\bar\psi_2 +\theta\theta m
        +\bar\theta\bar\theta n + \theta\sigma^m\bar\theta v_m
        +\theta\theta\bar\theta\bar\chi_1
        +\bar\theta\bar\theta\theta\chi_2
        +\theta\theta\bar\theta\bar\theta d,
\eea
where the action of the supersymmetry generators $Q$, $\bar Q$ on $f$ is
defined by $(\xi Q+\bar\xi\bar Q)\,f\equiv\delta_\xi f$. The explicit
transformation rules for the components of F are given below.

It is well known that the effect of a supersymmetry transformation acting 
on $F$ is a shift in superspace \cite{AV_tmp,WB}
\bea  \label{F_prime}
F'(x,\theta,\bar\theta) &= &e^{\xi Q+\bar\xi\bar Q}F(x,\theta,\bar\theta)
                                                    \nonumber\\
   &= &e^{(\theta+\xi)Q+(\bar\theta+\bar\xi)\bar Q
            +(\xi\sigma^m\bar\theta-\theta\sigma^m\bar\xi)P_m}f(x)
                                                    \nonumber\\
   &= &F(x^m-i(\xi\sigma^m\bar\theta-\theta\sigma^m\bar\xi),\theta+\xi,
         \bar\theta+\bar\xi)
       \ =\ F(x',\theta',\bar\theta').
\eea

The goldstino can be viewed as a hypersurface in superspace defined by
\be  \label{hypersurf}
\theta\ =\ -\kappa\,\lambda_g(x).
\ee
The requirement that this hypersurface be invariant under supersymmetry
transformations,
\be  \label{hypersurf_inv}
\theta'(x)\ =\ \theta(x')\qquad\Longrightarrow\qquad
-\kappa\,\lambda_g'(x)+\xi\ =\ -\kappa\,\lambda_g(x'),
\ee
implies (for infinitesimal $\xi$)
\be  \label{gstino_variation}
\lambda_g'(x)=\lambda_g(x)+{1\over\kappa}\xi
              +i\kappa\,(\xi\sigma^m\bar\lambda_g(x)
                 -\lambda_g(x)\sigma^m\bar\xi)\,
                 \partial_m\lambda_g(x),
\ee
which coincides with the transformation law (\ref{gstino_transf}).
The transformation of a field $\varphi(x)$ in the goldstino background
is determined by the condition that $\varphi(x)$ be well-defined on 
the hypersurface (\ref{hypersurf}), i.e., $\varphi'(x)=\varphi(x')$.
This implies the transformation law (\ref{matter_transf}). 

Now, consider the superfield $F$ restricted to the hypersurface
(\ref{hypersurf}),
\be  \label{f_hat}
\hat f(x)\ \equiv\ F(x,-\kappa\,\lambda_g(x),-\kappa\,\bar\lambda_g(x)).
\ee
Under an infinitesimal supersymmetry transformation it varies as
\be  \label{f_hat_variation}
\delta_\xi \hat f(x)\ =\ i\kappa\,(\xi\sigma^m\bar\lambda_g(x)
                 -\lambda_g(x)\sigma^m\bar\xi)\,\partial_m\hat f(x)
                    \ =\ -\kappa\,\vm\partial_m\hat f(x),
\ee
which is what we wanted to prove. It is straightforward to verify
explicitly that
\bea  \label{f_hat_expans}
\hat f &= &f -\kappa\,\lambda_g\psi_1 -\kappa\,\bar\lambda_g\bar\psi_2
        +\kappa^2\,\lambda_g\lambda_g m 
        +\kappa^2\,\bar\lambda_g\bar\lambda_g m
        +\kappa^2\,\lambda_g\sigma^m\bar\lambda_g v_m \nonumber\\
     && -\kappa^3\,\lambda_g\lambda_g\bar\lambda_g\bar\chi_1
        -\kappa^3\,\bar\lambda_g\bar\lambda_g\lambda_g\chi_2
        +\kappa^4\,\lambda_g\lambda_g\bar\lambda_g\bar\lambda_g d
\eea
varies as in (\ref{f_hat_variation}) using the component field variations
\bea  \label{F_comp_transf}
\delta_\xi f &= &\xi\psi_1+\bar\xi\bar\psi_2 \nonumber\\
\delta_\xi \psi_1 &=&2\,\xi m+\sigma^m\bar\xi\,(v_m+i\partial_m f) \nonumber\\
\delta_\xi \bar\psi_2 &=& 2\,\bar\xi n
                             +\bar\sigma^m\xi\,(-v_m+i\partial_m f)\nonumber\\
\delta_\xi v_m &=&\bar\chi_1\bar\sigma_m\xi-{i\over2}\,
                    \xi\sigma_n\bar\sigma_m\partial^n\psi_1
                   +\bar\xi\bar\sigma_m\chi_2+{i\over2}\,
                    \partial^n\bar\psi_2\bar\sigma_m\sigma_n\bar\xi\nonumber\\
\delta_\xi m &=& \bar\xi\bar\chi_1-{i\over2}\,\partial_m\psi_1\sigma^m\bar\xi
                   \nonumber\\
\delta_\xi n &=& \xi\chi_2+{i\over2}\,\xi\sigma^m\partial_m\bar\psi_2
                    \nonumber\\
\delta_\xi \bar\chi_1 &=&2\,\bar\xi d+{i\over2}\,\bar\sigma^n\sigma^m\bar\xi
                               \,\partial_m v_n
                             +i\,\bar\sigma^m\xi\,\partial_m m\nonumber\\
\delta_\xi \chi_2 &=&2\,\xi d-{i\over2}\,\sigma^n\bar\sigma^m\xi
                               \,\partial_m v_n
                             +i\,\sigma^m\bar\xi\,\partial_m n\nonumber\\
\delta_\xi d &=&{i\over2}\,\xi\sigma^m\partial_m\bar\chi_1
                -{i\over2}\,\partial_m\chi_2\sigma^m\bar\xi
\eea
and the goldstino variation (\ref{gstino_transf}).

It is now clear how to couple the components of the superfield $F$ to
bosonic or fermionic fields $\phi_i$ without superpartners in such 
a way that all interactions are invariant under the linear and 
non-linear supersymmetry transformations acting on $F$ and $\phi_i$
respectively.
Assume that the coupling of $f$ to the non-supersymmetric fields is 
of the form $\cP(f,\partial_m f,\phi_i,\partial_m \phi_i)$, where $\cP$ 
is some function of the fields and their derivatives. The supersymmetric
completion is then obtained by replacing partial derivatives by
covariant ones and $f$ by $\hat f$. That is, the invariant interaction
is $\cP(\hat f,D_m \hat f,\phi_i,D_m \phi_i)$ where $D_m$ is defined
in (\ref{cov_deriv}).

Let us now see how the chiral version of non-linear supersymmetry, eqs.\ 
(\ref{gstino_chiral}), (\ref{matter_chiral}), can be understood in 
this formalism. A chiral superfield $\Phi$ is a function of $x_L$ and
$\theta$ only, where 
\be  \label{xL_def}
x_L^m\ =\ x^m+i\theta\sigma^m\bar\theta.
\ee
In the $(x_L,\theta)$ basis, the superspace expansion of $\Phi$ is
simply given by
\be  \label{Phi_chiral_expans}
\Phi\ =\ e^{\theta Q}\phi\ =\ \phi + \sqrt2\,\theta\psi + \theta\theta F.
\ee
A supersymmetry transformation acting on $\Phi$ yields
\bea  \label{Phi_prime}
\Phi'(x_L,\theta) &= &e^{\xi Q+\bar\xi\bar Q}\Phi(x_L,\theta) \nonumber\\
         &= &e^{(\theta+\xi)Q-(2\,\theta\sigma^m\bar\xi\,
                               +\xi\sigma^m\bar\xi)\,P_m}
             e^{\bar\xi\bar Q}\phi(x_L)   \nonumber\\
         &= &\Phi(x_L^m+2i\theta\sigma^m\bar\xi+i\xi\sigma^m\bar\xi,
                  \theta+\xi)\ =\ \Phi(x_L',\theta'),
\eea
where we used $\bar Q_{\dot\alpha}\phi=0$.

The chiral goldstino $\tilde\lambda_g$ can be viewed as a hypersurface 
in chiral superspace defined by
\be  \label{chypersurf}
\theta\ =\ -\kappa\,\tilde\lambda_g(x_L).
\ee
The requirement that this hypersurface be invariant under supersymmetry
transformations,
\be  \label{chypersurf_inv}
\theta'(x_L)\ =\ \theta(x_L')\qquad\Longrightarrow\qquad
-\kappa\,\tilde\lambda_g'(x_L)+\xi\ =\ -\kappa\,\lambda_g(x_L'),
\ee
implies (for infinitesimal $\xi$)
\be  \label{chiral_gstino_variation}
\tilde\lambda_g'(x_L)=\tilde\lambda_g(x_L)+{1\over\kappa}\xi
              -2i\kappa\,\tilde\lambda_g(x_L)\sigma^m\bar\xi\,
                 \partial_m\tilde\lambda_g(x_L),
\ee
which coincides with the transformation law (\ref{gstino_chiral}).
As a byproduct, we find the relation \cite{SW}
$\tilde\lambda_g(x_L)=\lambda_g(x)$, where
$x_L^m=x^m+i\kappa^2\tilde\lambda_g(x_L)\sigma^m\bar{\tilde\lambda}_g(x_L)$.
The transformation of a field $\varphi(x_L)$ in the chiral goldstino background
is determined by the condition that $\varphi(x_L)$ be well-defined on 
the hypersurface (\ref{chypersurf}), i.e., $\varphi'(x_L)=\varphi(x_L')$.
This implies the transformation law (\ref{matter_chiral}). 

Now, consider the superfield $\Phi$ restricted to the hypersurface
(\ref{chypersurf}),
\be  \label{phi_hat_def}
\hat \phi_L(x_L)\ \equiv\ \Phi(x_L,-\kappa\,\tilde\lambda_g(x_L)).
\ee
This is exactly the composite field $\hat\phi_L$, defined in (\ref{phi_hat_L}).
Under an infinitesimal supersymmetry transformation it varies as
claimed above,
\be  \label{phi_hat_variation}
\delta_\xi \hat \phi_L(x_L)\ =\ -2i\kappa\,\tilde\lambda_g(x_L)\sigma^m\bar\xi
                                       \,\partial_m\hat \phi_L(x_L).
\ee

\subsection{Coupling a vector superfield to non-supersymmetric matter}
\label{Am_phi_coupl}
The case of a vector superfield is slightly more complicated because
the supersymmetry algebra only closes up to a gauge transformation. 
But we can learn some interesting physics by working out the couplings
of an $\cN=1$ vector to non-supersymmetric matter.\

Consider a vector superfield in the Wess-Zumino gauge
\be \label{V_expans}
V\ =\ -\theta\sigma^m\bar\theta A_m + i\,\theta\theta\bar\theta\bar\lambda
      - i\,\bar\theta\bar\theta\theta\lambda 
      + \hf\theta\theta\bar\theta\bar\theta D.
\ee
For simplicity we concentrate on an Abelian vector; the generalization
to non-Abelian gauge symmetry is straightforward.
The supersymmetry transformations are
\bea  \label{V_comp_transf}
\delta_\xi A_m &= &-i\,\lambda\sigma_m\bar\xi
                   +i\,\xi\sigma_m\bar\lambda, \nonumber\\
\delta_\xi \lambda &= &\sigma^{mn}\xi F_{mn} + i\,\xi D, \nonumber\\
\delta_\xi D &= &-\partial_m\lambda\sigma^m\bar\xi
                 -\xi\sigma^m\partial_m\bar\lambda.
\eea

The couplings of $A_m$ to bosonic fields $\phi_i$ are of the form
\be  \label{Am_phi}
\cL_0=-\sum_i\,(\partial^m +i\,q_iA^m)\phi_i\,
               (\partial_m -i\,q_iA_m)\phi_i^\dagger,
\ee
where $q_i$ are the charges of the fields $\phi_i$.
To find the supersymmetric completion of this Lagrangian, we first need to
find a superfield whose lowest component is $A_m$. This can be easily \
constructed by computing
\bea  \label{Am_superfield}
\cA_m &= &e^{\theta Q+\bar\theta\bar Q}A_m \nonumber\\
      &= &A_m + i\,\theta\sigma_m\bar\lambda - i\,\lambda\sigma_m\bar\theta
          -\theta\sigma^n\bar\theta\,(\tilde F_{mn}-\eta_{mn}D) \nonumber\\
     &&-\theta\theta\bar\theta\left(\bar\sigma_{mn}\partial^n\bar\lambda
                  -{1\over6}\partial_m\bar\lambda\right)
       -\bar\theta\bar\theta\theta\left(\sigma_{mn}\partial^n\lambda
                  -{1\over6}\partial_m\lambda\right)
       +\qt\theta\theta\bar\theta\bar\theta\,\partial^n F_{mn}.
\eea
From the previous subsection, we expect that the field that couples
covariantly to non-supersymmetric matter is
\bea  \label{Am_hat}
\hat A_m &= &A_m - i\kappa\,\lambda_g\sigma_m\bar\lambda 
              + i\kappa\,\lambda\sigma_m\bar\lambda_g
              - \kappa^2\,\lambda_g\sigma^n\bar\lambda_g\,
                                   (\tilde F_{mn}-\eta_{mn}D) \nonumber\\
         &&  + \kappa^3\,\lambda_g\lambda_g\bar\lambda_g\left(
                    \bar\sigma_{mn}\partial^n\bar\lambda
                    -{1\over6}\partial_m\bar\lambda\right)
             + \kappa^3\,\bar\lambda_g\bar\lambda_g\lambda_g\left(
                     \sigma_{mn}\partial^n\lambda
                    -{1\over6}\partial_m\lambda\right)   \nonumber\\
         &&  + \kappa^4\,\qt\lambda_g\lambda_g\bar\lambda_g\bar\lambda_g\,
                     \partial^n F_{mn}.
\eea
The explicit computation of the supersymmetry variation of $\hat A_m$
shows, however, that $\delta_\xi\hat A_m$ is not exactly of the form
(\ref{f_hat_variation}) but rather is given by
\be  \label{Am_hat_variation}
\delta_\xi \hat A_m\ =\ -\kappa\,\vn\,(\partial_n\hat A_m
                                      -\partial_m\hat A_n).
\ee
The additional second term is only at first sight unexpected. It can be 
traced to the fact that the commutator of two supersymmetry transformations 
acting on $A_m$ closes only up to a (field-dependent) gauge transformation,
\be \label{commmut_Am}
[\delta_\eta,\delta_\xi]\,A_m\ =\ 2i\,(\eta\sigma^n\bar\xi
                  -\xi\sigma^n\bar\eta)\,F_{mn}.
\ee
The term $\kappa\,\vn\,\partial_m\hat A_n$ in (\ref{Am_hat_variation})
is not a gauge transformation, but interestingly, we find that the
variation of $\hat A_m$ agrees with (\ref{gauge_transf}) up to a 
gauge transformation.
\be  \label{Am_hat_var}
\delta_\xi \hat A_m\ =\ -\kappa\,\vn\,\partial_n\hat A_m
                        -\kappa\,\partial_m\vn\,\hat A_n
                        +\kappa\,\partial_m(\vn\hat A_n).
\ee
This allows us to build gauge covariant derivatives defining
\be  \label{gaugecov_deriv}
\cD_m\phi_i\ =\ \left(\omega^{-1}\right)_m^{\ n}(\partial_n\phi_i
                   +i\,q_i\hat A_n\phi_i).
\ee
To cancel the gauge variation of $\hat A_m$ under $\delta_\xi$, we need
to modify the non-linear supersymmetry transformation of $\phi_i$
from the standard form (\ref{matter_transf}) to
\be  \label{charged_matter_transf}
\delta_\xi\phi_i\ =\ -\kappa\,\vm\,(\partial_m\phi_i+i\,q_i\hat A_m\phi_i).
\ee

A short calculation shows that the Lagrangian
\be  \label{Am_phi_susy}
\cL=-\det(\omega)\sum_i\cD^m\phi_i\left(\cD_m\phi_i\right)^\dagger
\ee
is indeed invariant up to a total derivative under the above 
supersymmetry transformations.

\section{Partially broken extended supersymmetry}
Let us see how the formalism presented in the preceding sections 
generalizes to the case of non-linearly realized $\cN=2$ supersymmetry.
We are interested in situations where an $\cN=1$ subgroup is still 
linearly realized, i.e., the extended supersymmetry is partially broken.
Specifically, we focus on cases where the spectrum and all interactions 
are manifestly $\cN=1$ supersymmetric but the $\cN=1$ superfields have 
no $\cN=2$ partners.
The goldstino $\lambda_g$ resides in an $\cN=1$ supermultiplet $\Lambda_g$
in such models. This can be a chiral \cite{BGc}, a vector \cite{BG} or a 
linear multiplet \cite{BGl}.
We concentrate on the case where the goldstino is the superpartner of a
$U(1)$ gauge boson because this seems to be most appropriate for the study
of supersymmetry breaking on D-branes.

\subsection{An action for the goldstino superfield}
We start by considering a model that only contains the goldstino and its
superpartner, a $U(1)$ gauge boson. The generalization of the goldstino 
transformation law (\ref{gstino_transf}) is \cite{BW}
\be  \label{sgstino_transf}
\delta^*_\eta\Lambda_g={1\over\kappa}\eta-\kappa\,\Vm\partial_m\Lambda_g,
\qquad{\rm where}\quad 
\Vm=i\Lambda_g\sigma^m\bar\eta-i\eta\sigma^m\bar\Lambda_g,
\ee
$\Lambda_g$ is a superfield with $\lambda_g$ as its lowest component
and the star on $\delta^*_\eta$ is to remind us that we are varying
with respect to the second supersymmetry. Again, it is useful to also
consider a chiral version of this transformation law \cite{BW}, which 
generalizes
(\ref{gstino_chiral}),
\be  \label{sgstino_chiral}
\delta^*_\eta\tilde\Lambda_g={1\over\kappa}\eta-2i\kappa\,
             \tilde\Lambda_g\sigma^m\bar\eta\,\partial_m\tilde\Lambda_g.
\ee
This transformation is related to the standard transformation 
(\ref{sgstino_transf}) through the field redefinition
$\Lambda_g(x^m,\theta,\bar\theta)\equiv\tilde\Lambda_g(x^m+i\kappa^2
  \Lambda_g\sigma^m\bar\Lambda_g,\theta,\bar\theta)$.

Since the goldstino superfield $\Lambda_{g\,\alpha}$, where $\alpha$
is a Weyl spinor index, has the same lowest component as the field
strength superfield $W_{\!g\,\alpha}$ associated to the $U(1)$ gauge boson,
it is natural to expect that one could identify these two superfields, 
$\Lambda_{g\,\alpha}\sim W_{\!g\,\alpha}$. This, however, can only be true
to zeroth order in an expansion in powers of $\kappa$. The reason is
that the transformation (\ref{sgstino_transf}) is not compatible with
the conditions that have to be satisfied by the field strength superfield:
\be  \label{Walpha_cond}
\bar D_{\dot\alpha}W_{\!g\,\beta}=0,\qquad 
D^\alpha W_{\!g\,\alpha}=\bar D_{\dot\alpha}\bar W_{\!g}^{\dot\alpha}.
\ee
Note that an identification of the form
$\tilde\Lambda_{g\,\alpha}\sim W_{\!g\,\alpha}$
is not possible either, because (\ref{sgstino_chiral}) is only 
compatible with the first but not with the second condition in
(\ref{Walpha_cond}).

If one insists that the goldstino is the lowest component of $W_{\!g\,\alpha}$,
one has to find a transformation law for $W_{\!g\,\alpha}$ which differs from
(\ref{sgstino_transf}) but still realizes the supersymmetry algebra and
is compatible with the conditions (\ref{Walpha_cond}). This transformation
was determined by the authors of \cite{BG} and is given by\footnote{Our
notation and conventions are summarized in the appendix. They differ from 
those of \cite{BG}.}
\be  \label{Walpha_transf}
\delta^*_\eta W_{\!g\,\alpha}\ =\ {2\over\kappa}\,\eta_\alpha 
                           +{\kappa\over2}\,\bar D^2\bar X\,\eta_\alpha
                   +2i\kappa\,(\sigma^m\bar\eta)_\alpha\,\partial_m X,
\ee
where $X$ is a chiral superfield (of mass dimension 3), determined through 
the recursive relation
\be  \label{X_def}
X\ =\ {\qt\, W_{\!g}^2\over 1+{\kappa^2\over4}\bar D^2\bar X}.
\ee
Using the tricks explained in \cite{BG}, this relation can be explicitly
solved for $X$. In our notation, the result is
\bea  \label{X_explicit}
X &= &\qt\,W_{\!g}^2\ -\ {\kappa^2\over32}\,\bar D^2
                     \left[{W_{\!g}^2\bar W_{\!g}^2\over
                          1-\hf\,A+\sqrt{1+\qt\,B^2-A}}\right],\nonumber\\
{\rm where} &&A\ =\ -{\kappa^2\over8}
                     \left(D^2W_{\!g}^2+\bar D^2\bar W_{\!g}^2\right),
                                                               \nonumber\\
            &&B\ =\ -{\kappa^2\over8}
                     \left(D^2W_{\!g}^2-\bar D^2\bar W_{\!g}^2\right).
\eea

To verify that (\ref{Walpha_transf}) realizes the supersymmetry algebra, 
note that one has \cite{BG}
\be  \label{X_transf}
\delta^*_\eta X\ =\ {1\over\kappa}\,W_{\!g}\eta.
\ee

An immediate consequence of the $X$ transformation law (\ref{X_transf}) 
is that
\bea  \label{X_Lagrangian}
\cL &= &\int d^2\theta\, X\ +\ \int d^2\bar\theta\,\bar X \nonumber\\
    &= &\qt\int d^2\theta\,W_{\!g}^2\ +\ 
        \qt\int d^2\bar\theta\,\bar W_{\!g}^2\ 
        +{\kappa^2\over8}\int d^2\theta d^2\bar\theta\,
         W_{\!g}^2 \bar W_{\!g}^2\ + \ \cO(\kappa^4)
\eea
is invariant up to a total derivative under the second supersymmetry.
This reduces to the usual Lagrangian for a supersymmetric gauge multiplet
in the limit $\kappa\to0$. The exciting feature of the Lagrangian
(\ref{X_Lagrangian}) is that its restriction to the bosonic terms 
coincides precisely with the Born-Infeld Lagrangian \cite{BI}. Indeed, 
it has been shown in \cite{BG} that the bosonic terms of (\ref{X_Lagrangian}) 
can be written as
\be  \label{BI_Lagr}
\cL_{\rm bos}\ =\ {1\over\kappa^2}\,
                  \left(1-\sqrt{-\det(\eta_{mn}+\kappa\,F_{mn})}\right).
\ee
The requirement of non-linearly realized $\cN=2$ supersymmetry implies
an $\cN=1$ supersymmetric generalization of the Born-Infeld action for
the goldstino superfield if we assume that the goldstino resides in a
vector multiplet.

We are now in a position to give the precise relationship between 
$W_{\!g}$ and $\tilde\Lambda_g$ that transforms according to the chiral 
standard transformation. One can verify that the superfield
\be  \label{Lambda_W_rel}
\tilde\Lambda_{g\,\alpha}\ \equiv\ {\hf\,W_{\!g\,\alpha}\over
                              1+{\kappa^2\over4}\bar D^2\bar X}
\ee
transforms as in (\ref{sgstino_chiral}). In terms of $\tilde\Lambda_g$,
the Lagrangian (\ref{X_Lagrangian}) can be written as
\bea  \label{Lambda_Lagr}
\cL &= &\int d^2\theta\,E_L\,\tilde\Lambda_g\tilde\Lambda_g\ +\ 
        \int d^2\bar\theta\,E_R\,\bar{\tilde\Lambda}_g\bar{\tilde\Lambda}_g,
                                            \nonumber\\
{\rm where} &&E_L=1+{\kappa^2\over4}\bar D^2\bar X,\qquad
              E_R=1+{\kappa^2\over4}D^2 X.
\eea
Note that $\tilde\Lambda_g$ and $E_L$ are chiral superfields,
i.e., $\bar D_{\dot\alpha}\tilde\Lambda_{g\,\beta}=0=\bar D_{\dot\alpha}E_L$.
From (\ref{X_transf}), one finds that $E_L$ transforms as a chiral density
under the non-linear supersymmetry transformation,
\be  \label{EL_transf}
\delta^\star_\eta E_L\ =\ -i\kappa\,\partial_mW\sigma^m\bar\eta\ 
     =\ -2i\kappa\,\partial_m(E_L\,\tilde\Lambda_g\sigma^m\bar\eta).
\ee

\subsection{$\cN=1$ matter fields in the goldstino background}
With the above results, it is straightforward to include $\cN=1$ matter
fields that couple to the goldstino in such a way that all interactions
are invariant under the non-linear $\cN=2$ supersymmetry. A chiral
superfield $\tilde\Phi$ transforms as \cite{BW}
\be  \label{Phi_transf}
\delta^\star_\eta\tilde\Phi\ =\ -2i\kappa\,\tilde\Lambda_g\sigma^m\bar\eta\,
                                         \partial_m\tilde\Phi.
\ee
From the transformation law (\ref{EL_transf}) of the chiral density $E_L$,
we find that 
\be  \label{Phi_Lagr}
\cL_{\rm pot}\ =\ \int d^2\theta\,E_L\,\cP(\tilde\Phi)\ +\ 
        \int d^2\bar\theta\,E_R\,\cP(\tilde\Phi^\dagger)
\ee
transforms into a total derivative for an arbitrary analytic function $\cP$.
This generalizes the leading order result obtained in \cite{BG} to all
orders in $\kappa$.

To find a generalization of the kinetic terms 
$\int d^2\theta d^2\bar\theta\,\Phi^\dagger\Phi$, we define
\bea  \label{shifted_fields}
\Lambda_g(x,\theta,\bar\theta) &= &\tilde\Lambda_g(x^m+i\kappa^2
     \Lambda_g\sigma^m\bar\Lambda_g,\theta,\bar\theta), \nonumber\\
\Phi(x,\theta,\bar\theta) &= &\tilde\Phi(x^m+i\kappa^2
     \Lambda_g\sigma^m\bar\Lambda_g,\theta,\bar\theta), \nonumber\\
E(x,\theta,\bar\theta) &= &E_L(x^m+i\kappa^2
     \Lambda_g\sigma^m\bar\Lambda_g,\theta,\bar\theta)
\eea
and note that $\Lambda_g$ transforms as the goldstino superfield 
(\ref{sgstino_transf}) and $\Phi$ transforms as
\be  \label{phi_real_transf}
\delta^\star_\eta\Phi\ =\ -i\kappa\,(\Lambda_g\sigma^m\bar\eta
         -\eta\sigma^m\bar\Lambda_g )\,\partial_m\Phi
      \ =\ -\kappa\,V^m_\eta\partial_m\Phi.
\ee
This is the standard non-linear realization for a general superfield.
A natural guess for an invariant action is 
\bea  \label{Kahler_Lagr}
\cL_{\rm kin} &= &\int d^2\theta d^2\bar\theta\,\hat E\,K(\Phi,\Phi^\dagger),
                                                    \nonumber\\
{\rm where} &&\hat E\ =\ \hf\,(E+E^\dagger)\ =\ 
           1+{\kappa^2\over8}\bar D^2\bar\Lambda_g^2
            +{\kappa^2\over8} D^2\Lambda_g^2+\cO(\kappa^4),
\eea
and the K\"ahler potential $K(\Phi,\Phi^\dagger)$ is an arbitrary real
function of $\Phi$, $\Phi^\dagger$. To order $\kappa^2$, this agrees with 
the Lagrangian of \cite{BG}. It is easy to check that the 
$\cO(1/\kappa)$ and the $\cO(\kappa)$ terms cancel in the variation
of the Lagrangian. Whether the invariance also holds to
higher orders in $\kappa$ is not clear to us.

A vector superfield $V$ can be chosen to transform according to the
standard non-linear realization (\ref{phi_real_transf}),
\be  \label{vector_transf}
\delta^\star_\eta V\ =\ -\kappa\,(i\Lambda_g\sigma^m\bar\eta-
         i\eta\sigma^m\bar\Lambda_g )\,\partial_m V.
\ee
The transformation of the field strength superfield $W_\alpha$ derived 
from the vector $V$ is non-standard and very complicated. But it is possible
to construct an $\cN=2$ covariant field strength $\cW_\alpha$ that
transforms according to (\ref{Phi_transf}).
This implies that 
\be  \label{gauge_coupl}
\cL_{\rm gauge}\ =\ \int d^2\theta d^2\bar\theta\,\hat E\,\Phi^\dagger e^V\Phi
                    \ +\ \qt\int d^2\theta\,E_L\,\cW\cW
                    \ +\ \qt\int d^2\bar\theta\,E_R\,\bar\cW\bar\cW
\ee
is invariant up to a total derivative.

To obtain $\cW_\alpha$, we have to introduce covariant derivatives
$\cD_\alpha$, $\bar\cD_{\dot\alpha}$, $D_m$ that reduce to $D_\alpha$,
$\bar D_{\dot\alpha}$, $\partial_m$ in the limit $\kappa\to0$ and satisfy
\be  \label{cov_super_deriv}
\delta^\star_\eta (\cD_\alpha V)\ =\ -\kappa\,V^m_\eta\partial_m
                                      (\cD_\alpha V)
\ee  \label{Ntwo_cov_deriv}
and similarly for $\bar\cD_{\dot\alpha}$, $D_m$. 
The explicit expressions for $\cD_\alpha$, $\bar D_{\dot\alpha}$, $D_m$, 
as derived by the authors of \cite{BG}, are given in the appendix. 
It follows that
\be  \label{W_zero}
\cW^0_\alpha\ =\ -\qt\,\bar\cD^2\cD_\alpha V
\ee
transforms according to the standard non-linear realization 
(\ref{phi_real_transf}). (For simplicity, we concentrate on an Abelian
gauge symmetry.) However, $\cW^0_\alpha$ is not gauge invariant.

To understand the reason for this, let us first see how gauge invariance
is achieved for the $\cN=1$ field strength superfield $W_\alpha=-\qt
\bar D^2 D_\alpha V$. A gauge transformation acts as 
$V\to V+i(\Lambda-\Lambda^\dagger)$ on the vector superfield, where
$\bar D_{\dot\alpha}\Lambda=0$. The latter condition ensures that a
chiral superfield $\Phi$ remains chiral under a gauge transformation
$\Phi\to e^{-i\Lambda}\Phi$. Gauge invariance of $W_\alpha$ follows
from the fact that $\bar D_{\dot\alpha}\Lambda=0$ by using the
commutation relation 
$[\bar D^2,D_\alpha]=4i(\sigma^m\bar D)_\alpha\partial_m$.

Returning to the case of non-linearly realized $\cN=2$ supersymmetry,
we note that the superfield $\Phi$ appearing in (\ref{gauge_coupl}) and
defined in (\ref{shifted_fields}) is not an $\cN=1$ chiral superfield
in the sense that $\bar D_{\dot\alpha}\Phi\neq0$. However, one can
show \cite{BG} that $\bar\cD_{\dot\alpha}\Phi=0$ if 
$\bar D_{\dot\alpha}\tilde\Phi=0$. Thus, the condition on the gauge
parameter $\Lambda$ has to be generalized to 
$\bar\cD_{\dot\alpha}\Lambda=0$. This guarantees that the $\cN=2$
chirality of $\Phi$, $\bar\cD_{\dot\alpha}\Phi=0$, is preserved
under gauge transformations.

An explicit computation of the commutation relation $[\bar\cD^2,\cD_\alpha]$
using the formulae given in the appendix yields
\bea  \label{cDcD_commut}
[\bar\cD^2,\cD_\alpha] &= &4i\,D_m\,(\sigma^m\bar\cD)_\alpha
   -4i\kappa^2\,\cD_\alpha(\Lambda_g\sigma^m)_{\dot\gamma}
    \bar\cD_{\dot\beta}\bar\Lambda_g^{\dot\gamma}\,D_m\bar\cD^{\dot\beta}
                                                   \nonumber\\
   &&-2i\kappa^2\,\cD_\alpha(\Lambda_g\sigma^m)_{\dot\gamma}\bar\cD^2
         \bar\Lambda_g^{\dot\gamma}\,D_m\ +\ \cO(\kappa^4).
\eea
This implies
\be   \label{cDcD_Lambda}
\bar\cD^2\cD_\alpha (\Lambda-\Lambda^\dagger)\ =\ 
       [\bar\cD^2,\cD_\alpha]\Lambda\ =\ 
       -2i\kappa^2\,\cD_\alpha(\Lambda_g\sigma^m)_{\dot\gamma}\bar\cD^2
         \bar\Lambda_g^{\dot\gamma}\,D_m\Lambda\ +\ \cO(\kappa^4).
\ee
As a consequence, $\cW^0_\alpha$ as defined in (\ref{W_zero}) is
not gauge invariant. But
\be  \label{cW_prime}
\cW_\alpha'\ =\ -\qt\left(\bar\cD^2\cD_\alpha-\kappa^2\,
         \cD_\alpha\Lambda_g^\beta\bar\cD^2
         \bar\Lambda_g^{\dot\gamma}\bar\cD_{\dot\gamma}\cD_\beta
                    +\cO(\kappa^4)\right)V
\ee
is gauge invariant and still transforms according to the standard 
non-linear realization (\ref{phi_real_transf}) up to order $\kappa^2$.
Finally,
\be  \label{cW_def}   
\cW_\alpha(x,\theta,\bar\theta)\ =\ 
\cW_\alpha'(x^m-i\kappa^2\,\Lambda_g\sigma^m\bar\Lambda_g,\theta,\bar\theta)
\ee
has the desired chiral transformation law (\ref{Phi_transf}).   

Note that it is not possible to couple charged matter to the goldstino
gauge multiplet. The term
\be  \label{gstino_gauge_coupl}
\cL_{\rm gauge}\ =\ \int d^2\theta d^2\bar\theta\,\hat E\,
                     \Phi^\dagger e^{V_g}\Phi
\ee
is not invariant, because of the shift in $V_g$. Here, $V_g$ is the
vector superfield containing $\lambda_g$, i.e., 
$W_{\!g\,\alpha}=-\qt\bar D^2D_\alpha V_g$. From the transformation
of $W_{\!g\,\alpha}$, eq.\ (\ref{Walpha_transf}), we find
\be  \label{gstino_vec_transf}
\delta^\star_\eta V_g\ =\ {2\over\kappa}\,
        \left(\bar\theta\bar\theta-\kappa^2\,\bar X\right)\theta\eta
        +{2\over\kappa}\,
        \left(\theta\theta-\kappa^2\,X\right)\bar\theta\bar\eta.
\ee
Therefore, $\delta^\star_\eta\cL_{\rm gauge}={2\over\kappa}
(\eta D+\bar\eta\bar D) (\hat E\,\Phi^\dagger e^{V_g}\Phi)
\hspace{0.5pt}\rule[-0.8ex]{0.3pt}{2.8ex}_{\theta=0}$.

This is not a problem for D-brane models because there is no matter
that is charged under the goldstino $U(1)$ in such models. To see this,
consider a stack of $N$ D-branes.
The gauge symmetry of the theory on world-volume of the D-branes is of
the form $\prod_i U(n_i)$, with $\sum_i n_i=N$. All matter fields are
in bifundamental representations, $(n_i,\overline{n_j})$, for some $i,j$
(this is the adjoint representation if $i=j$).
The goldstino $U(1)$ is diagonally embedded in the D-brane gauge group,
i.e., $U(1)_{\rm goldstino}=\sum_i U_i(1)$, where $U_i(1)$ is the trace
part of $U(n_i)$. As a consequence, all matter fields are neutral under
$U(1)_{\rm goldstino}$. This general argument applies to D-branes in
the bulk, which break $\cN=8$ supersymmetry to $\cN=4$, and to D-branes
at an orbifold fixed point breaking either $\cN=4$ supersymmetry to $\cN=2$
or $\cN=2$ supersymmetry to $\cN=1$. In the case of extended supersymmetry
on the D-branes, there is more than one goldstino. The world-volume
theory of a stack of $N$ D-branes in the bulk, for example, is an 
$N=4$ super-Yang-Mills theory with gauge group $SU(N)\times U(1)_{\rm
goldstino}$. The four goldstinos inside the $\cN=4$ $U(1)$ multiplet
correspond to the four broken supersymmetries.
The argument does not apply to D-branes at orientifold singularities. 
But supersymmetry breaking in orientifolds is not a partial breaking since 
bulk and (parallel) branes have the same amount of supersymmetry in such
models.

For the same reason, it is not possible to couple a chiral superfield
to the goldstino gauge kinetic terms. The Lagrangian
\be  \label{dil_gstino_coupl}
\cL\ =\ \int d^2\theta\,E_L\,\tilde\Phi\tilde\Lambda_g\tilde\Lambda_g\ +\ 
         \int d^2\bar\theta\,E_R\,\tilde\Phi^\dagger\bar{\tilde\Lambda}_g
         \bar{\tilde\Lambda}_g,
\ee
transforms as $\delta^\star_\eta\cL=
       {1\over\kappa}\int d^2\theta\,\tilde\Phi W_g\eta
      +{1\over\kappa}\int d^2\bar\theta\,\tilde\Phi^\dagger \bar W_g\bar\eta$,
which is not a total derivative.
This is puzzling because in string theory the gauge coupling constant is
related to the expectation value of the dilaton and the coupling of the 
dilaton superfield to the brane gauge fields should be of the form
(\ref{dil_gstino_coupl}).

\section{Coupling $\cN=2$ multiplets to $\cN=1$ matter}

\subsection{General formalism}
We now want to include fields on which $\cN=2$ supersymmetry is realized
linearly and determine their couplings to the $\cN=1$ superfields that
have no $\cN=2$ partners. The method will be a straightforward 
generalization of the one discussed in section 3. We work in the 
extended superspace spanned by the coordinates 
$(x,\theta,\bar\theta,\tilde\theta,\bar{\tilde\theta})$.
Let $\cF$ be an $\cN=2$ superfield\footnote{We restrict ourselves to
realizations of the $\cN=2$ supersymmetry algebra without central charge.
That is, we assume $\{Q_\alpha,S_\beta\}=0$.}
which has the $\cN=1$ superfield
$F$ as its lowest component in the $\tilde\theta$ expansion, i.e.,
\be  \label{cF}
\cF\ =\ e^{\tilde\theta S+\bar{\tilde\theta}\bar S}F,
\ee
where $S$, $\bar S$ are the generators of the second supersymmetry and 
the action of $S$, $\bar S$ on $F$ is defined by 
$(\eta S+\bar\eta\bar S)\,F\equiv\delta^\star_\eta F$.

The effect of a supersymmetry transformation acting on $\cF$ is a 
shift in extended superspace. For a transformation of the second 
supersymmetry, one has
\bea  \label{cF_prime}
\cF'(x,\theta,\bar\theta,\tilde\theta,\bar{\tilde\theta}) 
   &= &e^{\eta S+\bar\eta\bar S}
                 \cF(x,\theta,\bar\theta,\tilde\theta,\bar{\tilde\theta})
                                                    \nonumber\\
    &= &\cF(x^m-i(\eta\sigma^m\bar{\tilde\theta}
                  -\tilde\theta\sigma^m\bar\eta),\theta,\bar\theta,
                  \tilde\theta+\eta,\bar{\tilde\theta}+\bar\eta).
                                                    \nonumber\\
    &= &\cF(x',\theta',\bar\theta',\tilde\theta',\bar{\tilde\theta}{}')
\eea

The goldstino superfield can be viewed as a hypersurface in extended
superspace defined by
\be  \label{xhypersurf}
\tilde\theta\ =\ -\kappa\,\Lambda_g(x,\theta,\bar\theta).
\ee
The requirement that this hypersurface be invariant under transformations
of the second supersymmetry,
\be  \label{xhypersurf_inv}
\tilde\theta'(x,\theta,\bar\theta)\ =\ \tilde\theta(x',\theta',\bar\theta')
\qquad\Longrightarrow\qquad
-\kappa\,\Lambda_g'(x,\theta,\bar\theta)+\eta\ =\ -\kappa\,
         \Lambda_g(x',\theta',\bar\theta'),
\ee
implies (for infinitesimal $\eta$)
\be  \label{sgstino_variation}
\Lambda_g'(x,\theta,\bar\theta)=\Lambda_g(x,\theta,\bar\theta)
                                +{1\over\kappa}\eta
              +i\kappa\,(\eta\sigma^m\bar\Lambda_g(x,\theta,\bar\theta)
                 -\Lambda_g(x,\theta,\bar\theta)\sigma^m\bar\eta)\,
                 \partial_m\Lambda_g(x,\theta,\bar\theta),
\ee
which coincides with the transformation law (\ref{sgstino_transf}).

Now, consider the $\cN=2$ superfield $\cF$ restricted to the hypersurface
(\ref{xhypersurf}),
\be  \label{F_hat}
\hat F(x)\ \equiv \cF(x,\theta,\bar\theta,
                     -\kappa\,\Lambda_g(x,\theta,\bar\theta),
                     -\kappa\,\bar\Lambda_g(x,\theta,\bar\theta)).
\ee
Under an infinitesimal supersymmetry transformation it varies as
\bea  \label{F_hat_variation}
\delta^\star_\eta \hat F(x,\theta,\bar\theta) 
&= &i\kappa\,(\eta\sigma^m\bar\Lambda_g(x,\theta,\bar\theta)
               -\Lambda_g(x,\theta,\bar\theta)\sigma^m\bar\eta)\,
             \partial_m\hat F(x,\theta,\bar\theta) \nonumber\\
&= &-\kappa\,\Vm\partial_m\hat F(x,\theta,\bar\theta).
\eea
The composite superfield $\hat F$ reduces to $F$ in the limit $\kappa\to0$
and transforms according to the standard non-linear realization
(\ref{phi_real_transf}).

It is now clear how to couple the $\cN=1$ superfield $F$ that is the lowest
component of an $\cN=2$ superfield $\cF$ to $\cN=1$ superfields $\Phi_i$ 
that have no $\cN=2$ partners in such a way that all interactions are 
invariant under the full $\cN=2$ supersymmetry. If we know how $F$ couples
to the $\Phi_i$ in the limit $\kappa\to0$, then the $\cN=2$ completion
of these couplings is obtained by replacing $F$ by $\hat F$.

Let us now consider the chiral version of non-linear supersymmetry. 
A chiral $\cN=2$ superfield $\bfm\Phi$ is a function of $x_L$, $\theta$ 
and $\tilde\theta$ only, where 
\be  \label{xLtwo_def}
x_L^m\ =\ x^m+i\theta\sigma^m\bar\theta
             +i\tilde\theta\sigma^m\bar{\tilde\theta}.
\ee
In the $(x_L,\theta,\tilde\theta)$ basis, the $\tilde\theta$ expansion of 
$\bfm\Phi$ is simply given by
\be  \label{Phitwo_chiral_expans}
\bfm\Phi\ =\ e^{\tilde\theta S}\Phi\ =\ 
    \Phi + \sqrt2\,\tilde\theta\Psi + \tilde\theta\tilde\theta\bfm F,
\ee
where $\Phi$, $\Psi_\alpha$, $\bfm F$ are chiral $\cN=1$ superfields.

A supersymmetry transformation acting on $\bfm\Phi$ yields
\bea  \label{Phitwo_prime}
\bfm\Phi'(x_L,\theta,\tilde\theta) &= &e^{\eta S+\bar\eta\bar S}
                            \bfm\Phi(x_L,\theta,\tilde\theta) 
                                                         \nonumber\\
         &= &e^{(\tilde\theta+\eta)S-(2\,\tilde\theta\sigma^m\bar\eta\,
                               +\eta\sigma^m\bar\eta)\,P_m}
             e^{\bar\eta\bar S}\Phi(x_L,\theta)   \nonumber\\
         &= &\bfm\Phi(x_L^m+2i\tilde\theta\sigma^m\bar\eta
                           +i\eta\sigma^m\bar\eta,\theta,\tilde\theta+\eta)
              \ =\ \bfm\Phi(x_L',\theta',\tilde\theta'),
\eea
where we used that $\bar S_{\dot\alpha}\Phi=0$.

The chiral goldstino superfield $\tilde\Lambda_g$ can be viewed as a 
hypersurface in chiral extended superspace defined by
\be  \label{cxhypersurf}
\tilde\theta\ =\ -\kappa\,\tilde\Lambda_g(x_L,\theta).
\ee
The requirement that this hypersurface be invariant under supersymmetry
transformations,
\be  \label{cxhypersurf_inv}
\tilde\theta'(x_L,\theta)\ =\ \tilde\theta(x_L',\theta')
\qquad\Longrightarrow\qquad
-\kappa\,\tilde\Lambda_g'(x_L,\theta)+\eta\ =\ 
-\kappa\,\Lambda_g(x_L',\theta'),
\ee
implies (for infinitesimal $\eta$)
\be  \label{chiral_gstinotwo_variation}
\tilde\Lambda_g'(x_L,\theta)=\tilde\Lambda_g(x_L,\theta)+{1\over\kappa}\eta
              -2i\kappa\,\tilde\Lambda_g(x_L,\theta)\sigma^m\bar\eta\,
                 \partial_m\tilde\Lambda_g(x_L,\theta),
\ee
which coincides with the transformation law (\ref{sgstino_chiral}).
As a byproduct, we find the relation
$\tilde\Lambda_g(x_L,\theta)=\Lambda_g(x,\theta,\bar\theta)$, where
$x_L^m=x^m+i\theta\sigma^m\bar\theta
          +i\kappa^2\tilde\Lambda_g(x_L,\theta)\sigma^m
\bar{\tilde\Lambda}_g(x_L,\theta)$.
The transformation of an $\cN=1$ chiral superfield $\tilde\Phi(x_L,\theta)$ 
in the chiral goldstino background is determined by the condition that 
$\tilde\Phi(x_L,\theta)$ be well-defined on 
the hypersurface (\ref{cxhypersurf}), i.e., 
$\tilde\Phi'(x_L,\theta)=\tilde\Phi(x_L',\theta)$.
This implies the transformation law (\ref{Phi_transf}). 

Now, consider the $\cN=2$ superfield $\bfm\Phi$ restricted to the hypersurface
(\ref{cxhypersurf}),
\be  \label{Phitwo_hat_def}
\hat \Phi(x_L,\theta)\ \equiv\ \bfm\Phi(x_L,\theta,
             -\kappa\,\tilde\Lambda_g(x_L,\theta)).
\ee
Under an infinitesimal supersymmetry transformation it varies as
\be  \label{Phitwo_hat_variation}
\delta^\star_\eta \hat \Phi(x_L,\theta)\ =\ -2i\kappa\,
                   \tilde\Lambda_g(x_L,\theta)\sigma^m\bar\eta
                                       \,\partial_m\hat \Phi(x_L,\theta),
\ee
which is the chiral standard non-linear transformation law (\ref{Phi_transf}).

\subsection{Coupling an $\cN=2$ vector to $\cN=1$ multiplets}
Let us work out explicitly the couplings of an Abelian $\cN=2$ vector multiplet
to $\cN=1$ matter. An $\cN=2$ vector can be decomposed in an $\cN=1$
vector $V$ and an $\cN=1$ chiral multiplet $\Phi$.
\bea   \label{V_Phi}
V &= &-\theta\sigma^m\bar\theta A_m+i\theta\theta\bar\theta\bar\lambda^{(1)}
      -i\bar\theta\bar\theta\theta\lambda^{(1)}
      +\hf\,\theta\theta\bar\theta\bar\theta D \nonumber\\
\Phi &= &\phi+i\theta\sigma^m\bar\theta\,\partial_m\phi
          +\qt\,\theta\theta\bar\theta\bar\theta\,\Box\phi
          +\sqrt2\,\theta\lambda^{(2)}
          -{i\over\sqrt2}\,\theta\theta\partial_m\lambda^{(2)}
          \sigma^m\bar\theta+\theta\theta F
\eea
The field strength superfield $W_\alpha=-\qt\,\bar D^2D_\alpha V$ has
the component expansion
\bea  \label{Walpha_expans}
W_\alpha &= &-i\lambda^{(1)}_\alpha+\theta_\alpha D-i(\sigma^{mn}\theta)_\alpha
             F_{mn}+\theta\sigma^m\bar\theta\,\partial_m\lambda^{(1)}_\alpha
             +\theta\theta(\sigma^m\partial_m\bar\lambda^{(1)})_\alpha 
                                                                 \nonumber\\
         &&-\hf\,\theta\theta(\sigma^m\bar\theta)_\alpha
            (i\,\partial_m D-\partial^n F_{mn})
           -{i\over4}\,\theta\theta\bar\theta\bar\theta
                          \Box\lambda^{(1)}_\alpha.
\eea
The supersymmetry transformations are (see, e.g., \cite{West})
\bea  \label{vectortwo_transf}
\delta_\xi\phi &= &\sqrt2\left(\xi^{(1)}\lambda^{(2)}
                               -\xi^{(2)}\lambda^{(1)}\right),  \nonumber\\
\delta_\xi\lambda^{(1)} &= & \sigma^{mn}\xi^{(1)}F_{mn}+i\xi^{(1)}D
              -i\sqrt2\,\sigma^m\bar\xi^{(2)}\partial_m\phi
              -\sqrt2\,\xi^{(2)}F^\dagger,         \nonumber\\
\delta_\xi\lambda^{(2)} &= & \sigma^{mn}\xi^{(2)}F_{mn}-i\xi^{(2)}D
                  +i\sqrt2\,\sigma^m\bar\xi^{(1)}\partial_m\phi
                  +\sqrt2\,\xi^{(1)}F,             \nonumber\\
\delta_\xi A_m &= & i\xi^{(1)}\sigma_m\bar\lambda^{(1)} 
                    -i\lambda^{(1)}\sigma_m\bar\xi^{(1)}
                    +i\xi^{(2)}\sigma_m\bar\lambda^{(2)} 
                    -i\lambda^{(2)}\sigma_m\bar\xi^{(2)},\nonumber\\
\delta_\xi D &= & -\xi^{(1)}\sigma^m\partial_m\bar\lambda^{(1)} 
                  -\partial_m\lambda^{(1)}\sigma^m\bar\xi^{(1)}
                  +\xi^{(2)}\sigma^m\partial_m\bar\lambda^{(2)} 
                  +\partial_m\lambda^{(2)}\sigma^m\bar\xi^{(2)},\nonumber\\
\delta_\xi F &= & i\sqrt2
                  \left(\bar\xi^{(1)}\bar\sigma^m\partial_m\lambda^{(2)} 
                  +\partial_m\bar\lambda^{(1)}\bar\sigma^m\xi^{(2)}\right).
\eea
One finds that the supersymmetry algebra closes on $\Phi$, but on $V$ it
closes only up to a gauge transformation,
\bea  \label{susy_commut}
[\delta^\star_\eta,\delta^\star_\xi]\,\Phi &= &-2i(\eta\sigma^m\bar\xi
                         -\xi\sigma^m\bar\eta)\,\partial_m\Phi, \nonumber\\{}
[\delta^\star_\eta,\delta^\star_\xi]\,V &= &-2i(\eta\sigma^m\bar\xi
                         -\xi\sigma^m\bar\eta)\,\partial_m V
                +\theta\sigma^n\bar\theta\left(-2i(\eta\sigma^m\bar\xi
                         -\xi\sigma^m\bar\eta)\,\partial_n A_m\right).
\eea

We are interested in two different kinds of couplings. Firstly, the vector
superfield $V$ inside the $\cN=2$ vector can couple to $\cN=1$ chiral
superfields $\tilde\Phi_i$. Secondly, the chiral superfield $\Phi$
inside the $\cN=2$ vector can couple to $\cN=1$ chiral superfields
$\tilde\Phi_i$ and to $\cN=1$ field strength superfields $\tilde W_\alpha^a$.

Let us start by constructing the $\cN=2$ completion of the $\cN=1$
supersymmetric gauge coupling term
$\int d^2\theta d^2\bar\theta\,\tilde\Phi^\dagger e^V \tilde\Phi$.
According to the general formalism, developed above, we need to compute
the $\tilde\theta$ expansion of an $\cN=2$ superfield $\cV$ that has
the $\cN=1$ vector $V$ as its lowest component. Using
\bea  \label{vector_comp_transf}
\delta^\star_\eta V &= &-i\theta\sigma^m\bar\theta\,(\eta\sigma_m\bar D
             \Phi^\dagger+\bar\eta\bar\sigma_m D\Phi), \nonumber\\
\delta^\star_\eta \Phi &= &-i\sqrt2\,W\eta, \nonumber\\
\delta^\star_\eta W_\alpha &= &{i\over\sqrt2}\,\bar\eta\bar DD_\alpha\Phi
                  -{i\over2\sqrt2}\,\eta_\alpha\,\bar D^2\Phi^\dagger,
\eea
we find,
\bea  \label{calV_expans}
 \cV &= &e^{\tilde\theta S+\bar{\tilde\theta}\bar S}V \nonumber\\
     &= &V-\theta\sigma^m\bar\theta\bigg[
              i\left(\bar{\tilde\theta}\bar\sigma_m
              D\Phi+\tilde\theta\sigma_m\bar D\Phi^\dagger\right)
          -{1\over2\sqrt2}\,\tilde\theta\sigma^n\bar{\tilde\theta}
           \left( D\sigma_m\bar\sigma_nW+\bar D\bar\sigma_m\sigma_n
                  \bar W\right)                    \nonumber\\
     &&   -\tilde\theta\tilde\theta\bar{\tilde\theta}
           \left(\bar\sigma_{mn}\bar D\partial^n\Phi^\dagger
                  -{1\over6}\bar D\partial_m\Phi^\dagger\right)
          -\bar{\tilde\theta}\bar{\tilde\theta}\tilde\theta
           \left(\sigma_{mn}D\partial^n\Phi
                  -{1\over6}D\partial_m\Phi\right) \nonumber\\
     &&+{i\over4\sqrt2}\,\tilde\theta\tilde\theta
                         \bar{\tilde\theta}\bar{\tilde\theta}
              \left(D\sigma_{mn}\partial^nW-\bar D\bar\sigma_{mn}\partial^n
                    \bar W\right)\bigg].
\eea
Then, we define
\be  \label{V_hat_def}
\hat V\ \equiv\ \cV\hspace{0.5pt}\rule[-1.5ex]{0.3pt}{3.5ex}_{\tilde\theta
                                                     =-\kappa\,\Lambda_g}.
\ee

In the non-supersymmetric case, section \ref{Am_phi_coupl}, we saw that
the commutation relation $[\delta_\eta,\delta_\xi]=-2i(\eta\sigma^n\bar\xi
-\xi\sigma^n\bar\eta)F_{nm}$ implied the transformation law
$\delta_\xi\hat A_m=-\kappa\,v^n_\xi \hat F_{nm}$. It is easy to generalize
this to the present case. Taking into account the extra term in the
commutator $[\delta^\star_\eta,\delta^\star_\xi]\,V$, eq.\ (\ref{susy_commut}),
one finds that eq.\ (\ref{F_hat_variation}) is modified to
\be  \label{V_hat_transf}
\delta^\star_\eta \hat V\ =\ -\kappa\,V^m_\eta\,\partial_m\hat V
               -\kappa\,V^m_\eta\,\theta\sigma^n\bar\theta\,\partial_n
                \hat A_m,
\ee
where $\hat A_m$ is the coefficient of $-\theta\sigma^m\bar\theta$ 
in the superspace expansion of $\hat V$. This
coincides with the expression for $\hat A_m$ given in eq.\ (\ref{Am_hat}).

We want to couple $\hat V$ to $\cN=1$ chiral multiplets $\tilde\Phi_i$
that carry charges $q_i$ under the $U(1)$ gauge symmetry of the vector $V$.
To achieve this, we define
\be  \label{Phi_i_def}
\Phi_i(x,\theta,\bar\theta)\ =\ \tilde\Phi_i(x_L,\theta), \qquad
{\rm where}\ x_L^m=x^m+i\theta\sigma^m\bar\theta
                      +i\kappa^2\,\Lambda_g\sigma^m\bar\Lambda_g,
\ee
and modify the $\Phi_i$ transformation law (\ref{phi_real_transf}) to
\be  \label{phi_transf_gauge}
\delta^\star_\eta \Phi_i\ =\ -\kappa\,V^m_\eta\left(\partial_m\Phi_i
         -\hf q_i\,\theta\sigma^n\bar\theta\,\partial_n\hat A_m\Phi_i\right),
\ee
Knowing how $\hat V$ and $\Phi_i$ transform under the second supersymmetry,
we conclude that the Lagrangian
\bea  \label{V_Phi_coupl}
\cL &= &
\int d^2\theta d^2\bar\theta\,\hat E\,\Phi_i^\dagger e^{q_i\hat V} \Phi_i
 \nonumber\\
&= &\int d^2\theta d^2\bar\theta\,\left(1+i\kappa\,q_i\,
   \theta\sigma^m\bar\theta\,
  (\bar\Lambda_g\bar\sigma_m D\Phi+\Lambda_g\sigma_m\bar D\Phi^\dagger)\right)
  \tilde\Phi_i^\dagger e^{q_i V}\tilde\Phi_i
 \ +\ \cO(\kappa^2)
\eea
is invariant up to a total derivative.

Next, let us construct the $\cN=2$ completion of the $\cN=1$ supersymmetric
couplings $\int d^2\theta\,\left(\Phi\,\tilde W^a\tilde W^a
+\cP(\Phi,\tilde\Phi_i)\right)$, where $\Phi$ is the $\cN=1$ chiral
superfield inside the $\cN=2$ vector and $\cP$ is an arbitrary analytic
function. The $\tilde\theta$ expansion of the $\cN=2$ superfield $\bfm\Phi$
that has $\Phi$ as its lowest component is much simpler than the 
$\tilde\theta$ expansion of $\cV$. Since $\bar S_{\dot\alpha}\Phi=0$,
$\bfm\Phi$ is an $\cN=2$ chiral superfield. One finds
\be  \label{boldPhi_expans}
\bfm\Phi\ =\ \Phi - i\sqrt2\,\tilde\theta W
                  - \qt\tilde\theta\tilde\theta\bar D^2\Phi^\dagger.
\ee
This implies that
\be  \label{Phi_hat_def}
\hat\Phi\ \equiv\ \Phi + i\sqrt2\,\kappa\,\tilde\Lambda_g W
   - \qt\kappa^2\,\tilde\Lambda_g\tilde\Lambda_g\bar D^2\Phi^\dagger
\ee
transforms according to the chiral standard non-linear realization
(\ref{Phi_transf}) and
\be  \label{Phi_W_coupl}
\cL\ =\ \int d^2\theta\,E_L\,\left(\hat\Phi\tilde\cW^a\tilde\cW^a
                          +\cP(\hat\Phi,\tilde\Phi_i)\right)\ +\ h.c.
\ee
is invariant up to a total derivative.

\section{Pseudo-Supersymmetry}
The formalism explained in the previous sections is well-suited to
analyze a class of very interesting supersymmetry breaking scenarios
that arise naturally in string theory. In models containing D-branes
and anti-D-branes (e.g., \cite{bot_up}) or intersecting D-branes
(e.g., \cite{quasi}), different sectors of the theory break different
halves of supersymmetry, thus breaking supersymmetry completely but in
a non-local way. All broken supersymmetries are non-linearly realized.
Such a scenario is called \pseudo.\footnote{A related situation with
non-linearly realized $\cN=2$ supergravity is analyzed in \cite{BFKQ}.}

Let us consider a toy model containing three sectors, a bulk sector
with $\cN=2$ supersymmetry and two boundary sectors preserving
the first and second of the two bulk supersymmetries, respectively.
We concentrate on cases where there is only an $\cN=2$ vector 
$(A_m,\lambda^{(1)},\lambda^{(2)},\phi)$ in the bulk
that couples to chiral multiplets $\tilde\Phi^{(1)}$, $\tilde\Phi^{(2)}$
and gauge field strengths $\tilde W^{(1)}$, $\tilde W^{(2)}$ on the
boundaries. Here $\tilde\Phi^{(1)}$, $\tilde W^{(1)}$ are chiral
superfields with respect to the first supersymmetry and $\tilde\Phi^{(2)}$, 
$\tilde W^{(2)}$ are chiral superfields with respect to the second
supersymmetry.
\bea   \label{Phi_Phi_W_W}
\tilde\Phi^{(1)} &= &\phi^{(1)}+\sqrt2\,\theta\psi^{(1)}+\theta\theta F^{(1)}
                                              \nonumber\\
\tilde W^{(1)}_\alpha &= &-i\tilde\lambda^{(1)}_\alpha +\theta_\alpha D^{(1)}
         -i(\sigma^{mn}\theta)_\alpha F^{(1)}_{mn}
         +\theta\theta(\sigma^m\partial_m\bar{\tilde\lambda}{}^{(1)})_\alpha
                                              \nonumber\\ 
\tilde\Phi^{(2)} &= &\phi^{(2)}+\sqrt2\,\tilde\theta\psi^{(2)}
                               +\tilde\theta\tilde\theta F^{(2)}
                                              \nonumber\\
\tilde W^{(2)}_\alpha &= &-i\tilde\lambda^{(2)}_\alpha 
         +\tilde\theta_\alpha D^{(2)}
         -i(\sigma^{mn}\tilde\theta)_\alpha F^{(2)}_{mn}
         +\tilde\theta\tilde\theta(\sigma^m\partial_m\bar{\tilde\lambda}
          {}^{(2)})_\alpha
\eea
The $\cN=2$ vector $(A_m,\lambda^{(1)},\lambda^{(2)},\phi)$ can be either
split into two $\cN=1$ superfields with respect to the first supersymmetry.
These are the vector $V$ and the chiral multiplet $\Phi$ defined in 
(\ref{V_Phi}).
\bea   \label{VPhi}
V &= &-\theta\sigma^m\bar\theta A_m+i\theta\theta\bar\theta\bar\lambda^{(1)}
      -i\bar\theta\bar\theta\theta\lambda^{(1)}
      +\hf\,\theta\theta\bar\theta\bar\theta D \nonumber\\
\Phi &= &\phi+i\theta\sigma^m\bar\theta\,\partial_m\phi
          +\qt\,\theta\theta\bar\theta\bar\theta\,\Box\phi
          +\sqrt2\,\theta\lambda^{(2)}
          -{i\over\sqrt2}\,\theta\theta\partial_m\lambda^{(2)}
          \sigma^m\bar\theta+\theta\theta F
\eea
Or we can split the $\cN=2$ vector into two $\cN=1$ superfields
with respect to the second supersymmetry.
\bea  \label{V_Phi_prime}
V' &= &-\tilde\theta\sigma^m\bar{\tilde\theta} A_m
      +i\tilde\theta\tilde\theta\bar{\tilde\theta}\bar\lambda^{(2)}
      -i\bar{\tilde\theta}\bar{\tilde\theta}\tilde\theta\lambda^{(2)}
      -\hf\,\tilde\theta\tilde\theta\bar{\tilde\theta}\bar{\tilde\theta} D 
                                                            \nonumber\\
\Phi' &= &\phi+i\tilde\theta\sigma^m\bar{\tilde\theta}\,\partial_m\phi
          +\qt\,\tilde\theta\tilde\theta\bar{\tilde\theta}\bar{\tilde\theta}
           \,\Box\phi
          -\sqrt2\,\tilde\theta\lambda^{(1)}
          +{i\over\sqrt2}\,\tilde\theta\tilde\theta\partial_m\lambda^{(1)}
          \sigma^m\bar{\tilde\theta}-\tilde\theta\tilde\theta F^\dagger
\eea

There are two goldstino superfields $\Lambda_g$ and $\Lambda_g'$ in 
this class of models. $\Lambda_g(x,\theta,\bar\theta)$ is associated with 
the breaking of the second supersymmetry and 
$\Lambda_g'(x,\tilde\theta,\bar{\tilde\theta})$ is associated with 
the breaking of the first supersymmetry. In complete analogy to the
partially broken supersymmetry discussed above, $\Lambda_g'$ is related
to the chiral superfield $\tilde\Lambda_g'$ and the field strength
superfield $W_g'$ through eqs.\ (\ref{shifted_fields}), (\ref{Lambda_W_rel}), 
(\ref{X_explicit}) after putting primes on all fields and replacing 
$\theta$ by $\tilde\theta$.

To build an $\cN=2$ invariant action, we construct chiral densities
$E_L^{(1)}$ and $E_L^{(2)}$ from $W_g$ and $W_g'$, respectively, as in 
(\ref{Lambda_Lagr}), we form real densities $\hat E^{(1)}$ and $\hat E^{(2)}$
from $\Lambda_g$ and $\Lambda_g'$, respectively, as in (\ref{Kahler_Lagr}), 
and we define $\Phi^{(1)}$, $\Phi^{(2)}$ from $\tilde \Phi^{(1)}$, 
$\tilde\Phi^{(2)}$ as in (\ref{Phi_i_def}). Furthermore, we define the
field strength superfields $\tilde\cW^{(1)}$ and $\tilde\cW^{(2)}$ that
transform covariantly under the second and first supersymmetry,
respectively, as in (\ref{cW_prime}), (\ref{cW_def}). 
Finally, we need to construct the composite superfields $\hat V$, $\hat\Phi$ 
as in (\ref{V_hat_def}), (\ref{calV_expans}), (\ref{Phi_hat_def}), and 
similarly for $\hat V'$, $\hat\Phi'$.
To obtain $\hat V'$, $\hat\Phi'$, one first computes the $\cN=2$ superfields
$\cV'$, $\bfm\Phi'$ that have $V'$, $\Phi'$ as their lowest component
in the $\theta$ expansion and then replaces $\theta\to-\kappa'\,\Lambda_g'$. 
Here, we assumed that the first supersymmetry is broken at a scale
$\kappa^{\prime-1/2}$ and that $\Lambda_g'$ transforms as in
(\ref{sgstino_transf}) but with $\kappa$ replaced by $\kappa'$.

It is now straightforward to write down an $\cN=2$ invariant Lagrangian
for our toy model:
\be   \label{the_Lagr}
\cL\ =\ \cL_1\ +\ \cL_2\ +\ \cL_{\rm bulk}, \nonumber\\
\ee
where
\bea   \label{pseudo_susy_Lagr}
\cL_1 &= &\int d^2\theta d^2\bar\theta\, \hat E^{(1)}\, \Phi^{(1)\dagger}
                  e^{\hat V} \Phi^{(1)}\nonumber\\
      &&+\ \int d^2\theta\,E_L^{(1)}\bigg(\left(a+b\,\hat\Phi\right)
                        \tilde\cW^{(1)}\tilde\cW^{(1)}
                        +\cP(\hat\Phi,\tilde\Phi^{(1)})
                        +\tilde\Lambda_g\tilde\Lambda_g\bigg)
                        \ +\ h.c.,\nonumber\\
\cL_2 &= &\int d^2\tilde\theta d^2\bar{\tilde\theta}\, \hat E^{(2)}\, 
                 \Phi^{(2)\dagger}e^{\hat V'} \Phi^{(2)}\nonumber\\
      &&+\ \int d^2\tilde\theta\,E_L^{(2)}\bigg(\left(a'+b'\,\hat\Phi'\right)
                        \tilde\cW^{(2)}\tilde\cW^{(2)}
                        +\cP'(\hat\Phi',\tilde\Phi^{(2)})
                        +\tilde\Lambda_g'\tilde\Lambda_g'\bigg)
                        \ +\ h.c.,\nonumber\\
\cL_{\rm bulk} &= &\int d^2\theta d^2\bar\theta\,\Phi^{\dagger}e^V \Phi
                   \ +\ \qt\int d^2\theta\,WW
                   \ +\ \qt\int d^2\bar\theta\,\bar W\bar W,
\eea
and $a$, $b$, $a'$, $b'$ are undetermined coupling constants.
A non-trivial superpotential $\cP$ or $\cP'$ is possible only if either
the boundary fields $\tilde\Phi^{(1)}$, $\tilde\Phi^{(2)}$ transform in 
real representations of the bulk gauge symmetry or if there are several
oppositely charged fields on each boundary.

It is interesting to note that the interactions 
(\ref{pseudo_susy_Lagr}) preserve a $U(1)_R$ subgroup of the $SU(2)_R$ 
symmetry that is present in the $\cN=2$ theory. Under this $U(1)_R$ 
symmetry the fields carry charges as shown in table \ref{R_charges}.
This implies that the superfields $V$, $\Phi$, $V'$, $\Phi'$, $\hat V$,
$\hat\Phi$, $\hat V'$, $\hat\Phi'$, $E_L^{(i)}$ and $\hat E^{(i)}$
all have R-charge 0, the superfields $\Lambda_g$, $\tilde\cW^{(1)}$,
$W$ have R-charge 1 and the superfields $\Lambda_g'$, $\tilde\cW^{(2)}$,
$W'$ have R-charge $-1$.

\begin{table}[th]
\renewcommand{\arraystretch}{1.5}
$$\begin{array}{|c|c|c|c|c|c|c|c|c|c|c|c|c|}
\hline
{\rm field} &\theta &\tilde\theta &d^2\theta &d^2\tilde\theta 
            &\lambda_g &\lambda_g'
            &\phi &A_m &\lambda^{(1)} &\lambda^{(2)}
            &\tilde\lambda^{(1)} &\tilde\lambda^{(2)}\\ \hline
\hbox{R-charge} &1 &-1 &-2 &2 &1 &-1 &0 &0 &1 &-1 &1 &-1    \\ \hline
\end{array}$$
\caption{R-charges of the Grassmann variables $\theta$, $\tilde\theta$,
the goldstinos $\lambda_g$, $\lambda_g'$, 
the components of the bulk vector multiplet ($\phi$, $A_m$, 
$\lambda^{(1)}$, $\lambda^{(2)}$) and the boundary gauginos
$\tilde\lambda^{(1)}$, $\tilde\lambda^{(2)}$.
\label{R_charges}}
\end{table}

This toy model is very similar to open string models of intersecting
D-branes. Consider three stacks of D-branes at angles that intersect 
each other. Choose one of the stacks, say the third, and call it the bulk 
sector. Imagine that an orbifold compactification breaks supersymmetry
down to $\cN=2$ on each of the three stacks of D-branes, when considered
separately. But only the first supersymmetry is preserved at the
intersection of first with the third stack, and only the second 
supersymmetry is preserved at the intersection of the second with the
third stack. These intersections are the boundary sectors of our toy
model. In models of intersecting branes, all gauge fields live in the
bulk and the chiral matter fields are confined to the intersections.
Thus $a=b=a'=b'=0$ in such models. 
  
Supersymmetry is completely broken in the toy model described by
the above Lagrangian because $\cL_1$ and $\cL_2$ preserve 
different halves of the bulk $\cN=2$ supersymmetry.
What makes this supersymmetry breaking scenario very interesting
is that any two of the three subsectors $\cL_1$, $\cL_2$, $\cL_{\rm bulk}$
are supersymmetric when considered separately from the third sector.
That is obvious for $\cL_1+\cL_{\rm bulk}$ and $\cL_2+\cL_{\rm bulk}$.
For $\cL_1+\cL_2$, one has to set all bulk fields to zero and substitute
$\tilde\theta\to\theta$ in $\cL_2$. As a consequence, only interactions
that involve fields from all three sectors can generate mass splittings
for the components of the supersymmetric multiplets. Such interactions
only arise at two loops \cite{quasi}. This implies that the supersymmetric
non-renormalization theorems still apply at one loop. Thus there are no
mass splittings and no quadratic divergences at one loop.
The scalar masses squared arising at two loops are expected to be 
$\sim(g/4\pi)^4\,M^2$, where $g$ is the gauge coupling of the bulk 
vector and $M$ is the cut-off scale of the effective field theory.%
\footnote{This reasoning is valid before integrating out the auxiliary
fields $D$, $F$ of the bulk vector multiplet. After integrating out $D$,
$F$, the Langrangian contains direct couplings between the two boundary
sectors (this has been found in a similar context in \cite{ADGT}). 
For our toy model, we implicitly assume that the two boundary sectors are
seperated by some extra dimension such that direct couplings between
the two boundaries are absent even after integrating out the auxiliary
fields (see, e.g., \cite{MP}).}
A non-vanishing vacuum energy oly arises at three loops. This is similar
to a two-site moose model studied in \cite{AHCG}.

The fermion masses are protected to all orders in perturbation theory
by the $U(1)_R$ symmetry mentioned above. Generically, this $U(1)_R$ is 
broken down to a discrete subgroup by a quantum anomaly, but this discrete 
subgroup suffices to forbid fermion mass terms (assuming they are not already
present at tree-level). However, the $U(1)_R$ is broken explicitly, when 
the model is coupled to gravity. This is because the supercurrents 
corresponding to the two supersymmetries do not have well-defined
R-charges in the goldstino background. The supercurrents are polynomials
in the fields and their first derivatives. It turns out that different 
terms in these polynomials have different R-charges. To see this, first
consider the sector $\cL_1+\cL_{\rm bulk}$. Note that the goldstino 
transformation law (\ref{sgstino_transf}) is only compatible with the 
charge assignments of table \ref{R_charges} if we assign R-charge 1 to 
the supersymmetry variation parameter $\eta$ whereas the vector transformation
law (\ref{vectortwo_transf}) is only compatible with the charge assignments of 
table \ref{R_charges} if we assign R-charge $-1$ to the supersymmetry 
variation parameter $\eta=\xi^{(2)}$. This apparent mismatch is not a 
problem as long as we deal with global supersymmetry.
However, supergravity corrections include a term 
$\Mpl^{-1}\,\psi_m^{(2)}S^{(2)m}$, where $\psi_m^{(2)}$ is the
second gravitino and $S^{(2)m}$ is the Noether current associated
with the second supersymmetry. From the definition of the supercurrent,
\be  \label{Sm_def}
\delta^\star_{\eta,\rm local}\cL_{\rm global}\ =\ 
S^{(2)m}\partial_m\eta+\bar S^{(2)m}\partial_m\bar \eta,
\ee
it is clear that $S^{(2)m}$ has no well-defined R-charge if $\eta$ has no 
well-defined R-charge.

Interestingly, there is still an unbroken $\IZ_4$-symmetry that protects
the fermions from acquiring masses. Under this $\IZ_4$, all fields except
$\Lambda_g$ and $\Lambda_g'$ have charges as in table \ref{R_charges}.
The charges of the goldstino superfields are flipped, i.e., $\Lambda_g$
has charge $-1$ and $\Lambda_g'$ has charge 1. This implies that the
Lagrangian (\ref{the_Lagr}) is $\IZ_4$-invariant. It is easy to see that
the supercurrents have well-defined $\IZ_4$-charges ($S^{(1)m}:-1$,
$S^{(2)m}:1$) and that all interactions, including the coupling to 
supergravity, are $\IZ_4$-invariant if we assign charge 1 to $\psi^{(1)}_m$
and charge $-1$ to $\psi^{(2)}_m$.

Naively, one would expect that the second gravitino acquires a mass
by eating the first goldstino. Indeed, the coupling of the the second
gravitino to the second supercurrent,
$\Mpl^{-1}\,\psi_m^{(2)}S^{(2)m}$, 
contains a term\footnote{$\kappa^{-1/2}$ is the supersymmetry 
breaking scale; this is not related to the gravitational coupling,
which is set by $\Mpl$.} 
$(\Mpl\,\kappa)^{-1}\psi_m^{(2)}\sigma^m\bar\lambda_g$. But it is not
possible to eliminate the goldstino degrees of freedom by shifting the
gravitino field appropriately (see, e.g., \cite{AB}),
$\psi_m^{(2)'}=\psi_m^{(2)}
+(2\,\partial_m\lambda_g+im\,\sigma_m\bar\lambda_g)/\sqrt6\,m$, 
with $m=(\Mpl\,\kappa)^{-1}$.
The reason is that there is no mass term for $\lambda_g$. The goldstino
remains massless even when coupled to supergravity.\footnote{Note that
a similar result was found for a ten-dimensional non-supersymmetric
type I string model \cite{DM}.} 
Another way to see this is by considering the sector $\cL_1+\cL_{\rm bulk}$,
which has unbroken $\cN=1$ supersymmetry. The goldstino must remain
massless because it is the superpartner of the massless $U(1)$ gauge boson.
This has the surprising consequence that the second gravitino, too, remains
massless. Similarly, one finds that the first gravitino does not acquire
a mass.
Higher loop diagrams involving fields from all three sectors invalidate this
argument and most probably will generate gravitino masses. Since the goldstinos
are neutral under the bulk gauge symmetry, possible diagrams contributing to
goldstino masses only arise at three loops. The first contribution to the 
vacuum energy appears at the same loop order. This leads to a vacuum 
expectation value for one of the auxiliary fields of the gravity multiplet,
which cancels the vacuum energy term. This vacuum expectation value breaks
the above-mentioned $\IZ_4$-symmetry and generates gravitino masses.

\section{Conclusions and outlook}
We have developed a formalism to study effective field theory descriptions
of \pseudo. This is a supersymmetry breaking mechanism that
naturally arises in many open string models. Typically, such models contain
a bulk sector with extended supersymmetry and boundary sectors that break
different fractions of the bulk supersymmetry. But the complete supersymmetry
is still non-linearly realized. We have shown how to consistently couple
$\cN=2$ supersymmetric multiplets in the bulk to $\cN=1$ matter on the
boundaries. For a \pseudo\ toy model containing an $\cN=2$
vector in the bulk, we have computed explicitly the goldstino couplings.
One interesting result is that gravitino masses arise only at three loops.

There are several interesting directions for further research. First, it
is a straightforward exercise to apply the formalism of this article to
concrete D-brane models and to compute explicitly the scalar masses
arising at two loops.

Second, it is important to understand the coupling to supergravity in 
more detail. This was only sketched very roughly in the previous section. 
Most probably, one would gain new insights by generalizing the partial 
supergravity breaking of \cite{AB} to local \pseudo\ \cite{BFKQ}. 
It is also interesting to analyze the mechanism that breaks the 
$\IZ_4$-symmetry and generates gravitino masses. 

Finally, the coupling of bulk hyper multiplets to boundary gauge multiplets
is not discussed in this paper. It would certainly be very interesting
to determine these couplings. For a hyper multiplet $\bfm\Phi=(\Phi_1,\Phi_2)$,
where $\Phi_1$, $\Phi_2$ are chiral multiplets, it is easy to construct a
composite superfield $\hat\Phi_1$ that transforms linearly under the first
and non-linearly under the second supersymmetry. However, $\hat\Phi_1$ is
not a chiral superfield and therefore cannot be consistently coupled to
the gauge-kinetic terms of the boundary sector. It is well-known \cite{WLP}
that in linear $\cN=2$ supersymmetry, it is not possible to couple
hyper multiplets to the gauge kinetic terms of vector multiplets. The results
of the previous sections seem to suggest that it is not even possible to
couple components of hyper multiplets to gauge kinetic terms if the second
supersymmetry is non-linearly realized. This leads to a puzzle.
In $\cN=2$ supersymmetric type II string vacua, the dilaton resides in
a hyper multiplet. If there are D-branes, then the dilaton couples to 
the gauge-kinetic terms on their world-volume. It is not clear how this
coupling can be rendered non-linearly $\cN=2$ invariant.

\vskip1cm
\centerline{\bf Acknowledgements}

It is a pleasure to thank Michael Peskin for many stimulating discussions
and for proofreading the manuscript.  
My interest in pseudo-supersymmetry was strongly influenced by discussions
with Cliff Burgess, Elise Filotas and Fernando Quevedo.
I have also benefited from discussions
with Ignatios Antoniadis, Shamit Kachru and Albion Lawrence.
Special thanks go to my wife for her support and encouragement.
This research is funded by the Deutsche Forschungsgemeinschaft.

\vskip1cm

\begin{appendix}
\section{Notation and useful formulae}
\subsection{Spinors}
We use the metric $\eta_{mn}=\diag(-1,1,1,1)$ and the spinor
conventions of Wess and Bagger \cite{WB}. All spinors appearing in 
this article are 2-component Weyl spinors. A dot on an spinor index
indicates that the corresponding field transforms according to the
conjugate spinor representation of the Lorentz group.
Indices are raised and lowered with the
help of the antisymmetric tensors $\eps^{\alpha\beta}$, 
$\eps_{\alpha\beta}$ ($\eps^{12}=\eps_{21}=1$):
\bea  \label{ind_raise}
&& \psi^\alpha\ =\ \eps^{\alpha\beta}\psi_\beta,\quad
   \psi_\alpha\ =\ \eps_{\alpha\beta}\psi^\beta,\nonumber\\
&& \bar\psi^{\dot\alpha}\ =\ 
       \eps^{\dot\alpha\dot\beta}\bar\psi_{\dot\beta},\quad
   \bar\psi_{\dot\alpha}\ =\ \eps_{\dot\alpha\dot\beta}\bar\psi^{\dot\beta},
                                                                 \nonumber\\
&& \bar\sigma^{m\,\dot\alpha\alpha}\ =\ \eps^{\dot\alpha\dot\beta}
          \eps^{\alpha\beta}\sigma^m_{\beta\dot\beta}.
\eea
When spinor indices are suppressed, they are contracted as follows:
\be  \label{ind_contr}
\psi\chi\ =\ \psi^\alpha\chi_\alpha,\quad
\bar\psi\bar\chi\ =\ \bar\psi_{\dot\alpha}\bar\chi^{\dot\alpha},\quad
\psi\sigma^m\bar\chi\ =\ \psi^\alpha\sigma^m_{\alpha\dot\alpha}
                         \bar\chi^{\dot\alpha}.
\ee
Complex conjugation acts as
\be  \label{compl_conj}
(\psi_\alpha)^\dagger\ =\ \bar\psi_{\dot\alpha},\quad
(\psi\chi)^\dagger\ =\ \bar\chi\bar\psi,\quad
(\psi\sigma^m\bar\chi)^\dagger\ =\ \chi\sigma^m\bar\psi.
\ee
Some useful identities for the $\sigma$-matrices are
\bea  \label{sigma_ident}
(\sigma^{mn})_\alpha^{\ \beta} &\equiv &\qt\,(\sigma^m\bar\sigma^n
                      -\sigma^n\bar\sigma^m)_\alpha^{\ \beta},\\
(\sigma^m\bar\sigma^n+\sigma^n\bar\sigma^m)_\alpha^{\ \beta} &= &
                      -2\,\eta^{mn}\,\delta_\alpha^{\ \beta},\\
\sigma^m_{\alpha\dot\alpha}\bar\sigma_m^{\dot\beta\beta} &= &
             -2\,\delta_\alpha^{\ \beta}\,\delta_{\dot\alpha}^{\ \dot\beta},\\
{i\over2}\,\eps^{mnpq}\,\sigma_{pq} &= &\sigma^{mn}.
\eea
Some useful Weyl spinor identities are
\bea  \label{spinor_ident}
\theta^\alpha\theta^\beta &= &-\hf\,\eps^{\alpha\beta}\,\theta\theta,\quad
       \bar\theta_{\dot\alpha}\bar\theta_{\dot\beta}\ = \ 
        -\hf\,\eps_{\dot\alpha\dot\beta}\,\bar\theta\bar\theta,\\
\theta_\alpha\theta_\beta &= &\hf\,\eps_{\alpha\beta}\,\theta\theta,\quad
       \bar\theta^{\dot\alpha}\bar\theta^{\dot\beta}\ = \ 
        \hf\,\eps^{\dot\alpha\dot\beta}\,\bar\theta\bar\theta,\\
\psi\chi &= &\chi\psi,\\
\psi\sigma^m\bar\chi &= &-\bar\chi\bar\sigma^m\psi,\\
\psi\sigma^{mn}\chi &= &\chi\sigma^{nm}\psi,\\
(\theta\psi)(\chi\eta) &= &-\hf\,\Big[(\theta\eta)(\chi\psi)+
         (\theta\sigma^{mn}\eta)(\chi\sigma_{mn}\psi)\Big],\\
(\theta\psi)(\bar\chi\bar\eta) &= &\hf\,(\theta\sigma^m\bar\eta)
                                (\bar\chi\bar\sigma_m\psi).
\eea

\subsection{Supersymmetry}
The $\cN=2$ supersymmetry algebra without central charges is
\be  \label{N_two_susy}
\{ Q_\alpha,\bar Q_{\dot\beta}\}\ =\ \{S_\alpha,\bar S_{\dot\beta}\}\ =\ 
2\,\sigma^m_{\alpha\dot\beta}\,P_m,
\ee
with all other anticommutators vanishing. When acting on fields,
$P_m=-i\partial_m$. The $\cN=1$ covariant derivatives $D_\alpha$,
$\bar D_{\dot\alpha}$ are represented by differential operators
on superspace,
\be  \label{D_def}
D_\alpha\ =\ {\partial\over\partial\theta^\alpha}
             +i\,(\sigma^m\bar\theta)_\alpha\,\partial_m,\quad
\bar D_{\dot\alpha}\ =\ -{\partial\over\partial\bar\theta^{\dot\alpha}}
             -i\,(\theta\sigma^m)_{\dot\alpha}\,\partial_m.
\ee
They satisfy the anticommutation relations
\be  \label{D_algebra}
\{D_\alpha,D_\beta\}\ =\ 0,\qquad
\{D_\alpha,\bar D_{\dot\beta}\}\ =\ 
   -2i\,\sigma^m_{\alpha\dot\beta}\,\partial_m.
\ee
From (\ref{D_def}), we find
\bea  \label{D_squared}
D^2 &\equiv &D^\alpha D_\alpha\ =\ -{\partial\over\partial\theta}
          {\partial\over\partial\theta} -2i\bar\theta\bar\sigma^m
          {\partial\over\partial\theta}\partial_m -\bar\theta\bar\theta\Box,
                      \nonumber\\
\bar D^2 &\equiv &\bar D_{\dot\alpha} D^{\dot\alpha}\ =\ 
           -{\partial\over\partial\bar\theta}{\partial\over\partial\bar\theta} 
           -2i\theta\sigma^m{\partial\over\partial\bar\theta}\partial_m
           -\theta\theta\Box.
\eea
Integration of Grassmann variables and derivation with respect to them is
essentially equivalent. Normalizing the Grassmann integral measure such
that $\int d^2\theta\,\theta\theta=1$, one has
\be  \label{int_deriv}
\int d^2\theta\ =\ -\qt\,D^2\ +\ {\rm total\ deriv.},\quad
\int d^2\bar\theta\ =\ -\qt\,\bar D^2\ +\ {\rm total\ deriv.}
\ee

A chiral superfield $\Phi(x,\theta,\bar\theta)$ is defined by the condition
$\bar D_{\dot\alpha}\Phi=0$. Its complex conjugate is an antichiral
superfield, i.e., $D_\alpha\Phi^\dagger=0$.
For a chiral superfield
\be  \label{chiral_sf}
\Phi\ =\ \phi+i\theta\sigma^m\bar\theta\,\partial_m\phi
          +\qt\,\theta\theta\bar\theta\bar\theta\,\Box\phi
          +\sqrt2\,\theta\psi-{i\over\sqrt2}\,\theta\theta\,\partial_m\psi
          \sigma^m\bar\theta+\theta\theta F,
\ee
one has
\be  \label{D_sq_Phi}
-\qt\,D^2\Phi\ =\ F+i\sqrt2\,\bar\theta\bar\sigma^m\partial_m\psi
      -i\theta\sigma^m\bar\theta\,\partial_m F +\bar\theta\bar\theta\Box\phi
      +{1\over\sqrt2}\,\bar\theta\bar\theta\theta\Box\psi
      +\qt\,\theta\theta\bar\theta\bar\theta\,\Box F,
\ee
which is an antichiral superfield, since $D_\alpha D_\beta D_\gamma\equiv0$.
For the $\cN=1$ chiral field strength superfield
\bea  \label{Walpha_def}
W_\alpha &= &-i\lambda_\alpha+\theta_\alpha D-i(\sigma^{mn}\theta)_\alpha
             F_{mn}+\theta\sigma^m\bar\theta\,\partial_m\lambda_\alpha
             +\theta\theta(\sigma^m\partial_m\bar\lambda)_\alpha \nonumber\\
         &&-\hf\,\theta\theta(\sigma^m\bar\theta)_\alpha
            (i\,\partial_m D-\partial^n F_{mn})
           -{i\over4}\,\theta\theta\bar\theta\bar\theta\Box\lambda_\alpha
\eea
this yields
\be  \label{D_sq_W}
-\qt\,D^2 W\eta\ =\ -i\,\eta\sigma^m\partial_m\bar W.
\ee

The $\cN=2$ covariant derivatives introduced on page \pageref{Ntwo_cov_deriv}
are given by \cite{BG}
\bea  \label{cD_def}
\cD_\alpha &= &D_\alpha+i\kappa^2\left(D_\alpha\Lambda_g\sigma^m\bar\Lambda_g
                +D_\alpha\bar\Lambda_g\bar\sigma^m\Lambda_g\right)D_m,
                               \nonumber\\
\bar\cD_{\dot\alpha} &= &\bar D_{\dot\alpha}+i\kappa^2\left(
                 \bar D_{\dot\alpha}\Lambda_g\sigma^m\bar\Lambda_g
                +\bar D_{\dot\alpha}\bar\Lambda_g\bar\sigma^m\Lambda_g
                              \right)D_m,\nonumber\\
D_m &= &\left(\omega^{-1}\right)_m^{\ n}\partial_n,
\eea
where $\omega_m^{\ n}$ is a generalization of (\ref{omega_def}) to the
case of partially broken $\cN=2$ supersymmetry,
\be  \label{omeg_def}
\omega_m^{\ n}=\delta_m^{\ n}
               -i\kappa^2\left(\partial_m\Lambda_g\sigma^n\bar\Lambda_g
                  +\partial_m\bar\Lambda_g\bar\sigma^n\Lambda_g\right).
\ee
These are derivatives with respect to the first (linearly realized) 
supersymmetry but they are covariant with respect to the second 
(non-linearly realized) supersymmetry in the sense that they satisfy
(\ref{cov_super_deriv}). They generate the following algebra \cite{BG}
\bea  \label{cD_algebra}
\{\cD_\alpha,\cD_\beta\} &= & 2i\kappa^2\left(\cD_\alpha
          (\Lambda_g\sigma^m)_{\dot\gamma}\cD_\beta\bar\Lambda_g^{\dot\gamma}
          +\cD_\beta(\Lambda_g\sigma^m)_{\dot\gamma}\cD_\alpha
            \bar\Lambda_g^{\dot\gamma}\right)D_m,   \nonumber\\
\{\cD_\alpha,\bar\cD_{\dot\beta}\} &= &-2i\,\sigma^m_{\alpha\dot\beta}D_m
            +2i\kappa^2\left(\cD_\alpha(\Lambda_g\sigma^m)_{\dot\gamma}
                 \bar\cD_{\dot\beta}\bar\Lambda_g^{\dot\gamma}
          +\bar\cD_{\dot\beta}(\Lambda_g\sigma^m)_{\dot\gamma}\cD_\alpha
            \bar\Lambda_g^{\dot\gamma}\right)D_m,
            \nonumber\\{}
[\cD_\alpha,D_m] &= &2i\kappa^2\left(\cD_\alpha
           (\Lambda_g\sigma^n)_{\dot\gamma}D_m\bar\Lambda_g^{\dot\gamma}
          +D_m(\Lambda_g\sigma^n)_{\dot\gamma}\cD_\alpha
            \bar\Lambda_g^{\dot\gamma}\right)D_n.
\eea
Due to the fact that $\tilde\Lambda_g$ is an $\cN=1$ chiral superfield,
which is equivalent to the condition
\be  \label{cD_Lambda}
\bar\cD_{\dot\alpha}\Lambda_g\ =\ 0,
\ee
the algebra (\ref{cD_algebra}) simplifies to
\bea  \label{cD_algebra_simpl}
\{\cD_\alpha,\cD_\beta\} &= &0,\nonumber\\
\{\cD_\alpha,\bar\cD_{\dot\beta}\} &= &-2i\,\sigma^m_{\alpha\dot\beta}D_m
            +2i\kappa^2\,\cD_\alpha(\Lambda_g\sigma^m)_{\dot\gamma}
                 \bar\cD_{\dot\beta}\bar\Lambda_g^{\dot\gamma}\,D_m,
            \nonumber\\{}
[\cD_\alpha,D_m] &= &2i\kappa^2\,\cD_\alpha
           (\Lambda_g\sigma^n)_{\dot\gamma}D_m\bar\Lambda_g^{\dot\gamma}\,D_n.
\eea

\end{appendix}


\begin{thebibliography}{99}

\bibitem{Polch} J.~Polchinski,
                {\em ``Dirichlet Branes And Ramond-Ramond Charges''},
                \PhysRL{75} (1995) 4724, hep-th/9510017. 

\bibitem{ADS} I.~Antoniadis, E.~Dudas, A.~Sagnotti,
              {\em ``Brane Supersymmetry Breaking''},
              \PhysL{464} (1999) 38, hep-th/9908023.

\bibitem{AU} G.~Aldazabal, A.~M.~Uranga, 
             {\em ``Tachyon Free Nonsupersymmetric Type IIB 
                    Orientifolds Via Brane - Anti-Brane Systems''},
             JHEP 9910 (1999) 024, hep-th/9908072. 

\bibitem{AIQ} G.~Aldazabal, L.~E.~Ib\'a\~nez, F.~Quevedo,
              {\em ``Standard - Like Models With Broken Supersymmetry 
                     From Type I String Vacua''},
              JHEP 0001 (2000) 031, hep-th/9909172.

\bibitem{AADDS} C.~Angelantonj, I.~Antoniadis, G.~D'Appollonio, E.~Dudas, 
                A.~Sagnotti,
                {\em ``Type I vacua with brane supersymmetry breaking''},
                \Nucl{572} (2000) 36, hep-th/9911081.

\bibitem{BGKL} R.~Blumenhagen, L.~G\"orlich, B.~K\"ors, D.~L\"ust,
                {\em ``Noncommutative Compactifications Of Type I Strings 
                       On Tori With Magnetic Background Flux},
                JHEP 0010 (2000) 006, hep-th/0007024;
                R.~Blumenhagen, B.~K\"ors, D.~L\"ust, T.~Ott,
                {\em ``The Standard Model From Stable Intersecting Brane 
                       World Orbifolds''},
                \Nucl{616} (2001) 3, hep-th/0107138. 

\bibitem{quasi} D.~Cremades, L.~E.~Ib\'a\~nez, F.~Marchesano,
                {\em ``SUSY Quivers, Intersecting Branes and the Modest 
                       Hierarchy Problem''},
                hep-th/0201205; 
                {\em ``Intersecting Brane Models of Particle Physics and 
                       the Higgs Mechanism''},
                hep-th/0203160.

\bibitem{string} J.~Polchinski,
                {\em ``String Theory''}, Ch.~13.2, 
                Cambridge University Press (1998).

\bibitem{BG} J.~Bagger, A.~Galperin, 
              {\em ``New Goldstone multiplet for partially 
                     broken supersymmetry''},
              \PhysR{55} (1997) 1091, hep-th/9608177. 

\bibitem{BFKQ} C.~Burgess, E.~Filotas, M.~Klein, F.~Quevedo,
               work in progress.

\bibitem{MP} E.~A.~Mirabelli, M.~E.~Peskin, 
             {\em ``Transmission Of Supersymmetry Breaking From A 
                    Four-Dimensional Boundary''},
             \PhysR{58} (1998) 065002, hep-th/9712214. 

\bibitem{DM} E.~Dudas, J.~Mourad,
             {\em ``Consistent gravitino couplings in 
                    non-supersymmetric strings''},
             \PhysL{514} (2001) 173, hep-th/0012071.

\bibitem{PR} G.~Pradisi, F.~Riccioni,
             {\em ``Geometric Couplings and Brane Supersymmetry Breaking},
             \Nucl{615} (2001) 33, hep-th/0107090.

\bibitem{ABL} I.~Antoniadis, K.~Benakli, A.~Laugier,
              {\em ``D-brane Models with Non-Linear Supersymmetry''},
              hep-th/0111209.

\bibitem{AV} D.~V.~Volkov, V.~P.~Akulov,
             {\em ``Possible Universal Neutrino Interaction''},
             {\it JETP~ Lett.} {\bf 16} (1972) 438;
             {\em ``Is The Neutrino A Goldstone Particle?''},
             \PhysL{46} (1973) 109.

\bibitem{AV_tmp} V.~P.~Akulov, D.~V.~Volkov,
             {\em ``Goldstone Fields With Spin 1/2''},
             {\it Theor.~Math.~Phys.} {\bf 18} (1974) 28.
 
\bibitem{Pash} A.~I.~Pashnev,
              {\em ``Nonlinear Realization For Symmetry Group With
                     Spinor Parameters''},
              {\it Theor.~Math.~Phys.} {\bf 20} (1974) 725.

\bibitem{VS} D.~V.~Volkov, V.~A.~Soroka,
             {\em ``Higgs Effect For Goldstone Particles With Spin 1/2''}
             {\it JETP~Lett.} {\bf 18} (1973) 312.

\bibitem{WZ} J.~Wess, B.~Zumino, 
             {\em ``Supergauge Transformations In Four-Dimensions''},
             \Nucl{70} (1974) 39. 

\bibitem{WB} J.~Wess, J.~Bagger,
             {\em ``Supersymmetry and Supergravity''}, 2nd ed., 
             Princeton University Press (1992).

\bibitem{IK} E.~A.~Ivanov, A.~A.~Kapustnikov,
             {\em ``General relationship between linear and nonlinear 
                    realisations of supersymmetry''},
             {\it J.~Phys.} {\bf A11} (1978) 2375.

\bibitem{CL} T.~E.~Clark, S.~T.~Love,
             {\em ``Goldstino Couplings to Matter''},
              \PhysR{54} (1996) 5723, hep-ph/9608243.

\bibitem{CLLW} T.~E.~Clark, T.~Lee, S.~T.~Love, G.~H.~Wu,
               {\em ``On the Interactions of Light Gravitinos''},
               \PhysR{57} (1998) 5912, hep-ph/9712353.

\bibitem{Z} B.~Zumino,
            {\em ``Fermi-Bose Supersymmetry (Supergauge Symmetry In 
                   Four-Dimensions)''},
            Proc.\ of 17th Int.\ Conf.\ on High Energy Physics, 
            Imperial College, London, England, Jul 1-10, 1974,
            J.~R.~Smith, ed., Rutherford Lab.\ (1974) I-254.

\bibitem{SW} S.~Samuel, J.~Wess,
             {\em ``A superfield formulation of the nonlinear realization 
                    of supersymmetry and its coupling to supergravity''},
             \Nucl{221} (1983) 153.

\bibitem{BGc} J.~Bagger, A.~Galperin,
              {\em ``Matter couplings in partially broken extended 
                     supersymmetry''},
              \PhysL{336} (1994) 25, hep-th/9406217. 

\bibitem{BGl} J.~Bagger, A.~Galperin, 
              {\em ``The tensor Goldstone multiplet for partially 
                     broken supersymmetry''},
              \PhysL{412} (1997) 296, hep-th/9707061. 

\bibitem{BW} J.~Bagger, J.~Wess,
             {\em ``Partial breaking of extended supersymmetry''},
             \PhysL{138} (1984) 105.

\bibitem{BI} M.~Born, L.~Infeld,
             {\em ``Foundations of the new field theory''},
             {\it Proc.~Roy.~Soc.~Lond.} {\bf A144} (1934) 425;
             R.~G.~Leigh,
             {\em ``Dirac-Born-Infeld action from Dirichlet sigma model''},
             {\it Mod.~Phys.~Lett.} {\bf A4} (1989) 2767.
             
\bibitem{West} P.~West,
               {\em ``Introduction to supersymmetry and supergravity''},
               World Scientific (1986).

\bibitem{bot_up} G.~Aldazabal, L.~E.~Ib\'a\~nez, F.~Quevedo, A.~M.~Uranga,
                 {\em ``D-Branes at Singularities: A Bottom-Up Approach to 
                        the String Embedding of the Standard Model''},
                 JHEP 0008 (2000) 002, hep-th/0005067.

\bibitem{ADGT} A.~Anisimov, M.~Dine, M.~Graesser, S.~Thomas,
               {\em ``Brane World Susy Breaking From String / M Theory''},
               hep-th/0201256.

\bibitem{AHCG} N.~Arkani-Hamed, A.~G.~Cohen, H.~Georgi,
               {\em ``Twisted Supersymmetry And The Topology Of 
                      Theory Space''},
               hep-th/0109082.

\bibitem{AB} R.~Altendorfer, J.~Bagger,
             {\em ``Dual Supersymmetry Algebras from Partial 
                    Supersymmetry Breaking''},
             \PhysL{460} (1999) 127, hep-th/9904213.

\bibitem{WLP} B.~de~Wit, P.~G.~Lauwers, A.~Van~Proeyen,
              {\em ``Lagrangians Of N=2 Supergravity - Matter Systems''},
              \Nucl{255} (1985) 569. 

\end{thebibliography}
\end{document}